\documentclass[twocolumn,showpacs,amsmath]{revtex4}
\usepackage{graphicx,amsmath,amsfonts, amssymb}
 \usepackage{color}
 \usepackage{xcolor}

\newcommand{\STO}{SrTiO$_3$}
\newcommand{\LVO}{LaVO$_3$}
\newcommand{\KNO}{KNbO$_3$}

\begin{document}

\title{First-principles Modelling of \STO\ based Oxides for Thermoelectric Applications}

\author{Daniel I. Bilc, Calin G. Floare, Liviu P. Z\^arbo, Sorina Garabagiu}
\affiliation{Mol $\&$ Biomol Phys Dept, Natl Inst Res $\&$ Dev Isotop$\&$ Mol Technol, RO-400293 Cluj-Napoca}
\author{Sebastien Lemal, and Philippe Ghosez}
\affiliation{Physique Th\'eorique des Mat\'eriaux, Q-MAT, CESAM, Universit\'e de Li\`ege (B5), B-4000 Li\`ege, Belgium}


\begin{abstract}

Using first-principles electronic structure calculations, we studied the electronic and thermoelectric properties of \STO\ based oxide materials and their nanostructures identifying those nanostructures which possess highly anisotropic electronic bands. We showed recently that highly anisotropic flat-and-dispersive bands can maximize the thermoelectric power factor, and at the same time they can produce low dimensional electronic transport in bulk semiconductors. Although most of the considered nanostructures show such highly anisotropic bands, their predicted thermoelectric performance is not improved over that of \STO. Besides highly anisotropic character, we emphasize the importance of the large weights of electronic states participating in transport and the small effective mass of charge carriers along the transport direction. These requirements may be better achieved in binary transition metal oxides than in ABO$_3$ perovskite oxide materials.

\end{abstract}

\maketitle

\section{INTRODUCTION}

Thermoelectric (TE) technology exploits the ability of certain materials for direct and reversible conversion of thermal energy into electricity. This double edge ability gives TE technology a strong appeal in almost all energy-related applications. Nevertheless, many fundamental problems have to be solved and in particular the intrinsic efficiency of TE materials still needs to be significantly improved before TE technology becomes a competitive alternative. The efficiency of a TE material depends on the dimensionless figure of merit, $ZT=(S^2\sigma$T$)/\kappa_{th}$, where $\sigma$ is the electrical conductivity, $S$ is the thermopower or Seebeck coefficient, $T$ is the absolute temperature,  $\kappa_{th}$ is the total thermal conductivity including electronic and lattice contributions, and $S^2\sigma$ is the power factor ($PF$). Improving TE efficiency is not obvious because the parameters entering $ZT$ are interlinked, and cannot be optimized independently. Moreover, $ZT$ has to be optimized in a $T$ range at which the TE devices will operate. Many interesting TE applications function at high $T$, and for practical applications other aspects are also very important such as: the cost of materials, their stability at high $T$, their toxicity, and availability.

Oxide materials are not known to exhibit the highest TE performance, but they offer stability in oxidizing and corrosive environments at high $T$ (T$ > $800 - 1000 K).   
In this context, oxides appear very appealing for high $T$ applications. Consequently, efforts have been devoted over the last 15 years to the optimization of TE properties of both n-type and p-type oxide materials. Good candidates for n-type materials include Nb-, W-, La-, Ce-, Pr-, Nd-, Sm-, Gd-, Dy-, and Y-doped SrTiO$_3$~\cite{Ohta2005, Okuda, Zhu2011, Mei2011, Mei2013, Hyeon2014, Weidenkaff2014, Buscaglia2014, Koumoto2015, Wu2015, Tritt2015,  Reaney2016, Chen2016}, Nb-, La-, Nd-, Sm-, and Gd-doped Sr$_2$TiO$_4$, and Sr$_3$Ti$_2$O$_7$~\cite{Koumoto2009}, La-doped CaMnO$_3$~\cite{Matsubara} or Al-, Ge-, Ni- and Co-doped ZnO~\cite{Ohtaki, Arai1997, Zeng2013, Ichinose2015, Zhai2015}, Ce-doped In$_2$O$_3$~\cite{Yang2015}, Er-doped CdO~\cite{Fu2013, Wang2015}, TiO$_2$~\cite{Brown2014}, Nb$_2$O$_5$~\cite{Brown2014}, WO$_3$~\cite{Brown2014}, while for p-type materials the most promising compounds are Ca$_3$Co$_4$O$_9$~\cite{Masset} with $ZT$$\sim$ 0.3 at 1000 K~\cite{Ohta2007}, and BiCuSeO with $ZT$$\sim$ 1.4 at 923 K.\cite{Zhao2013} Many studies have concerned doped \STO, demonstrating the largest $ZT$$\sim$ 0.4 in SrTi$_{0.8}$Nb$_{0.2}$O$_{3}$ films at 1000 K~\cite{Okuda}, and $ZT$$\sim$ 0.41 in bulk Sr$_{1 - 3x/2}$La$_x$TiO$_{3}$ at 973 K.\cite{Reaney2016}  Different strategies have been proposed to further increase TE efficiency.  Attempts to decrease the lattice thermal conductivity $\kappa_{l}$ by atomic substitution of Sr by Ba have been envisaged but seem to negatively affect the TE performance.\cite{Ohta2008} A more promising approach is the reduction of  $\kappa_{l}$ from scattering of phonons at interfaces, highlighted in layered Ruddlesden-Popper (RP) compounds.~\cite{Koumoto}  Also, Ohta {\it et. al.}  demonstrated significant enhancement of $S$ arising from electron confinement and the formation of two dimensional electron gas (2-DEG) in \STO/Nb-\STO\ superlattices.\cite{Ohta2007} Although very promising, the inactive \STO\ interlayer must be sufficiently large to avoid electron tunneling and to get significant enhancement of $S$~\cite{Mune}, but this decreases the effective TE performance. In order to maximize TE performance, we showed that the 2-DEG has to be achieved in doped semiconducting nanostructures rather than those with metallic character.\cite{Garcia} Few first-principles studies have investigated the electronic properties of Nb-doped \STO ~\cite{Astala, Guo, Zhang} and Sr$_2$TiO$_4$~\cite{Yun} and a few studies have addressed TE properties of  \STO ~\cite{Usui, Zhang2, Kinaci, Ricinschi2013, Zou2013, Yamanaka2013, Kahaly2014, Singsoog2014, Zhang3}, BaTiO$_3$~\cite{Zhang3}, PbTiO$_3$~\cite{Roy2016}, CaTiO$_3$~\cite{Zhang3}, KTaO$_3$~\cite{Janotti2016}, HoMnO$_3$~\cite{Ahmad2015}, CaMnO$_3$~\cite{Zhang2011, Zhang4, Molinari2014, Zhang2015}, Ca$_3$Co$_4$O$_9$~\cite{Maensiri2016}, ZnO~\cite{Ong, Jia2011, Jantrasee2014, Chen2015, Zheng2015}, Cu$_2$O~\cite{Chen}, CdO~\cite{Parker2015}, TiO$_2$~\cite{Kioupakis2015}, and V$_2$O$_5$~\cite{Chumakov}. At this stage, a complete understanding of the transport properties and moreover of the band structure engineering in oxide materials is still missing.

Employing the concept of electronic band structure engineering and our guidance ideas, we showed recently that very anisotropic flat-and-dispersive electronic bands are able to maximze the $PF$ and carrier concentration $n$ of bulk semiconductors, and at the same time to produce low-dimensional electronic transport (low-DET).\cite{Bilc2015} In practice this is typically achieved from the highly directionsl character of some orbitals like the $d$ states. Transition metal (TM) oxides with $d$-type conduction states appear as a well suited playground to explore this concept. Also the rich crystal chemistry of oxides encourages strategies of multiscale nanostructuring~\cite{Biswas, Heremans} by considering hybrid crystal structures that contain discrete structural blocks or layers. The nanostructuring of such hybrid materials is possible in order to engineer their electronic band structures and to lower their $\kappa_{l}$. 
Therefore, in this theoretical work, we studied TE properties of \STO\  based materials and their nanostructures. We have considered the (\STO)$_m$-(\LVO)$_1$ and (\STO)$_m$-(\KNO)$_1$ superlattices (SL).  \LVO\ is a Mott insulator with an electronic band gap $E_{g}\sim$1.1 eV~\cite{Inaba}, which shows a $PF$$\sim$ 0.06 $m  W/mK^2$ at 1000 K. \cite{Wang} For (\STO)$_m$-\LVO\ nanostructures we expect an electronic transport achieved through \LVO\ layers since it has a smaller band gap that \STO\ ($E_{g}\sim$3.3 eV), which may give rise to low-DET and enhanced $PF$. We consider (\STO)$_m$-\KNO\  SL in order to compare their results with those of Nb-doped \STO. We also studied TE properties of Sr and Co based naturally-ordered Ruddlesden-Popper compounds (AO[ABO$_3$]$_m$), which are more easy to control experimentally and more realistic for practical applications than artificial nanostructures.

\section{TECHNICAL DETAILS}

 The structural, electronic and TE properties of considered oxide structures were studied within density functional theory (DFT) formalism using the hybrid functional B1-WC.\cite{B1-WC} B1-WC hybrid functional describes the electronic properties (band gaps) and the structural properties with a better accuracy than the usual simple functionals, being more appropriate for correlated materials such as oxides.\cite{B1-WC, Goffinet, Prikockyte} For comparison of bulk \STO\ results, we also used approximations based on LDA~\cite{LDA}, GGA(PBE~\cite{PBE} and GGA-WC~\cite{GGA-WC}) usual simple functionals. The electronic structure calculations have been performed using the linear combination of atomic orbitals method as implemented in CRYSTAL first-principles code.\cite{Crystal} We used localized Gaussian-type basis sets including polarization orbitals and considered all the electrons for Ti \cite{Bredow}, O \cite{Piskunov2004}, V~\cite{Mackrodt}, K and Nb~\cite{Dovesi1991}, F and Co~\cite{Peintinger}. The Hartree-Fock pseudopotential for Sr~\cite{Piskunov2004}, and the Stuttgart energy-consistent pseudopotential for La~\cite{Cao} were used. 

In order to go beyond the rigid band approximation, we considered 3$\times$3$\times$3 \STO\ perovskite supercells with $P \overline{\it 1}$ symmetry, which incorporate explicitly two Nb and two La doping elements per supercell. For (\STO)$_m$-(\LVO)$_1$ and (\STO)$_m$-(\KNO)$_1$ nanostructures, we have considered $a \times a \times c$ SL with $P4mm$ symmetry to treat the nonmagnetic and ferromagnetic (FM) order, and $\sqrt{2} a \times \sqrt{2} a \times c$ SL with $P4bm$ symmetry for the FM and antiferromagnetic (AFM) orders, where $a$ is the lattice constant of cubic perovskite structure. For AO[ABO$_3$]$_m$ SL with $I4/mmm$ symmetry the body centered primitive cells were used in calculations. According to the position of F at the apical site, the symmetry of Sr$_2$CoO$_3$F has been reduced to $Cmcm$ and $Cmmm$ space groups for type I and type II  ordered structures, respectively. For these F ordered structures, we used the face centered primitive cells in our calculations.

Brillouin zone integrations were performed using the following meshes of $k$-points: 6$\times$6$\times$6  for bulk \STO, Sr$_2$TiO$_4$, and Sr$_2$CoO$_3$F, 3$\times$3$\times$3  for  La and Nb doped 3$\times$3$\times$3 \STO\ supercells,  6$\times$6$\times$4  for (\STO)$_1$-(\LVO)$_1$ and (\STO)$_1$-(\KNO)$_1$ SL, 6$\times$6$\times$1  for (\STO)$_5$-(\LVO)$_1$ and (\STO)$_5$-(\KNO)$_1$ SL, and 4$\times$4$\times$4  for Sr$_3$Ti$_2$O$_7$. The self-consistent-field calculations were considered to be converged when the energy changes between interactions were smaller than 10$^{-8}$ Hartree. An extra-large predefined pruned grid consisting of 75 radial points and 974 angular points was used for the numerical integration of charge density.  Full optimizations of the lattice constants and atomic positions have been performed with the optimization convergence of 5$\times$10$^{-5}$ Hartree/Bohr in the root-mean square values of forces and 1.2$\times$10$^{-3}$ Bohr in the root-mean square values of atomic displacements. The level of accuracy in evaluating the Coulomb and exchange series is controlled by five parameters.\cite{Crystal} The values used in our calculations are 7, 7, 7, 7, and 14.  

The transport coefficients were estimated in the Boltzmann transport formalism and the constant relaxation time approximation using BoltzTraP  transport code.\cite{Boltztrap} The electronic band structures (energies), used in the transport calculations, were calculated with electronic charge densities converged for denser $k$-point meshes (doubling the $k$-point meshes used in optimization calculations).  The transport coefficients were very well converged for the energies calculated on $k$-point meshes of 59$\times$59$\times$59 for bulk \STO, Sr$_2$TiO$_4$, Sr$_3$Ti$_2$O$_7$, and Sr$_2$CoO$_3$F, 27$\times$27$\times$27 for  La doped 3$\times$3$\times$3 \STO\ supercell,  59$\times$59$\times$41 for (\STO)$_1$-(\LVO)$_1$ and (\STO)$_1$-(\KNO)$_1$ SL, and 59$\times$59$\times$23 for (\STO)$_5$-(\LVO)$_1$ and (\STO)$_5$-(\KNO)$_1$ SL.

The effective masses were obtained by calculating values of energy close to conduction band (CB) minimum and valence band (VB) maximum while moving from the extremum points along the three directions of the orthogonal reciprocal lattice vectors $k_{i}$ (i=x, y, z). The energy values, $\epsilon_{\vec{k}}$, were fitted up to 10-th order polynomials in $k_{i}$.
In general $\epsilon_{\vec{k}}$ can be expanded about an extremum point as:
\begin{equation}
\frac{2m_{e}}{\hbar^{2}} \epsilon_{\vec{k}} = \sum_{i,j} \frac{m_{e}}{m_{ij}} k_{i}k_{j} 
\end{equation}
where $m_{ij}$ are the components of the effective mass tensor (i, j=x, y, z) and $m_{e}$ is the free electron mass. In the present expansion, we have only the diagonal components of the effective mass tensor $m_{ii}$, since all the interaxial angles are 90$^\circ$.

\section{RESULTS}

The oxide compounds under study are well known to exhibit various types of structural phase transitions with temperature. Since we are interested in TE at high T, in first approximation we  restricted our investigation to the high symmetry phases of  $P m\overline{\it 3}m$ for bulk perovskites and $P4mm$, $P4bm$, and $I4/mmm$ for superlattices. Some of them are also expected to show magnetic order at low T and be in a paramagnetic configuration at high T. For the later we compare different magnetic orders to find the systems with lower total energy and more structurally stable. 

\subsection{Bulk \STO\ and its alloys}

The structural (lattice constant) and electronic (band gap) properties of bulk \STO\ optimized within the different functionals are given in Table~\ref{Table1}. LDA  underestimates  the lattice constant and the atomic volume, whereas PBE overestimates these properties, which is the typical behaviour of LDA and PBE functionals. GGA-WC describes very well the structural properties of \STO, being a functional developed for solids. All the simple functionals underestimate \STO\ band gap, which is an inherent problem of DFT. B1-WC hybrid, which mixes the GGA-WC with a small percentage of exact exchange (16$\%$) describes simultaneously both the structural and electronic properties with good accuracy.

\begin{table}[t]
\caption{\label{Table1}Lattice constant $a$, and indirect band gap $E_g$ of \STO\ perovskite structure estimated within different functionals. The experimental values are included for comparison.  }
 \begin{tabular*}{0.49\textwidth}%
    {@{\extracolsep{\fill}}cccccc}
\hline\hline
                & LDA & PBE & GGA-WC & B1-WC & Exp.  \\ 
\hline
$a$(\AA) & 3.864  &	3.943   &	3.898   &	3.880     &	3.890$^a$  \\
$E_g (eV)    $ & 2.24 &	2.25 &	2.25 &	3.57 &	3.25$^b$ \\
\hline\hline
\multicolumn{6}{l}{ $^a$Extrapolated at 0 K from Ref.~\cite{Hellwege}. } \\
\multicolumn{6}{l}{ $^b$From Ref.~\cite{Benthem}.  }\\
\end{tabular*}
\end{table}

\begin{figure}[t]
\centering\includegraphics[scale=0.3]{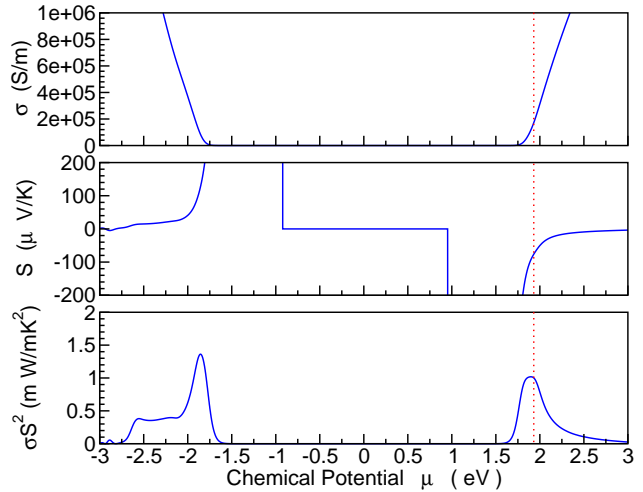}\\[-10pt]
\caption{\label{TranSTO} (Color online) Electrical conductivity $\sigma$, Seebeck coefficient $S$, and power factor $PF=S^2 \sigma$ dependence on chemical potential for \STO\ estimated at 300 K within B1-WC using the relaxation time $\tau=0.43\times10^{-14}$ s. The n-type doping carrier concentration $n$ of $1\times10^{21}$ cm$^{-3}$ is shown in dashed vertical line.}
\end{figure}

\begin{figure}[t]
\centering\includegraphics[scale=0.26]{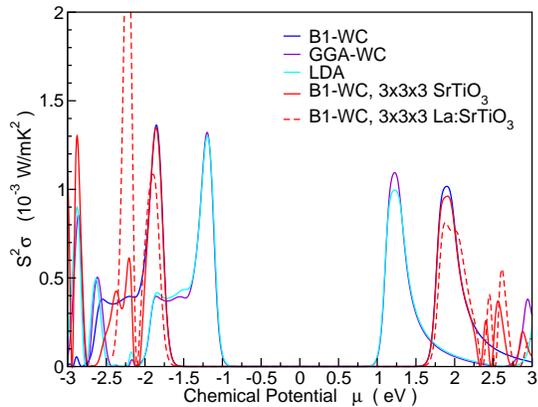}\\[-10pt]
\caption{\label{PFAllFunc} (Color online) Power factor $PF=S^2 \sigma$ dependence on chemical potential for \STO\ estimated at 300 K within different functionals using the relaxation time $\tau=0.43\times10^{-14}$ s. The different $PF$ peak positions within GGA-WC and LDA are due to lower band gap $E_g$ values. $PF$ for 3$\times$3$\times$3 \STO, and La:\STO\ supercells, estimated at 300 K within B1-WC, are also shown.}
\end{figure}

For a more complete theoretical characterization, we have studied also TE properties ($\sigma$, $S$, and $PF$) of bulk \STO\ within B1-WC hybrid, and the other used simple functionals.  In the constant relaxation time approximation, the relaxation time $\tau$ is considered as a constant $\tau=\tau_0$ independent of energy and temperature $T$, and is estimated from fitting of the experimental electrical conductivity $\sigma_{exp}$ at a given doping carrier concentration $n$ and $T$. $\tau$ within the different functionals was estimated by fitting the room temperature experimental values $\sigma_{exp}=1.4\times10^5$ S/m at $n=8\times10^{20}$ cm$^{-3}$~\cite{Muta}, and $\sigma_{exp}=1.667\times10^5$ S/m at $n=1\times10^{21}$ cm$^{-3}$~\cite{Okuda} (Table 2). The resulting value of $\tau$ is $\sim0.43\times10^{-14}$ s, which very similar in all the used functionals (see Table~\ref{Table2}).
The estimated values of  $\sigma$, $S$, and $PF$ at 300 K as a function of chemical potential within B1-WC hybrid are given in Fig.~\ref{TranSTO}.  For n-type doping, $PF$ is $\sim$1 mW/mK$^2$, being underestimated with respect to experiment ($PF_{exp} \sim$3 mW/mK$^2$).\cite{Muta, Okuda} This underestimation of $PF$ is due to a low value of thermopower  $S\sim$ -77 $\mu$V/K in comparison with the experimental value of $S_{exp} \sim$ -147 $\mu$V/K at  $n=1\times10^{21}$ cm$^{-3}$ and 300 K. $PF$ estimated within the usual simple functionals (LDA, GGA-WC) has the same value $\sim$1 mW/mK$^2$ (Fig.~\ref{PFAllFunc}).

\begin{table}[t]
\caption{\label{Table2}Relaxation time $\tau$ estimated at 300 K within constant relaxation time approximation using different functionals for n-type doped \STO.  }
 \begin{tabular*}{0.49\textwidth}%
    {@{\extracolsep{\fill}}cccc}
\hline\hline
                &$\quad$ LDA  $\quad$ & GGA-WC & B1-WC   \\ 
\hline
               & \multicolumn{3}{c}{$\sigma_{exp}=1.4\times10^5$ S/m, $n=8\times10^{20}$ cm$^{-3}$} \\ 
$\tau$( 10$^{-14}$ s) & 0.44 &	0.46 &  0.43  \\
               & \multicolumn{3}{c}{ $\sigma_{exp}=1.667\times10^5$ S/m, $n=1\times10^{21}$ cm$^{-3}$} \\ 
$\tau$( 10$^{-14}$ s) & 0.43 &	0.45 &  0.42  \\
\hline\hline
\end{tabular*}
\end{table}

\begin{figure}[t]
\centering\includegraphics[scale=0.23]{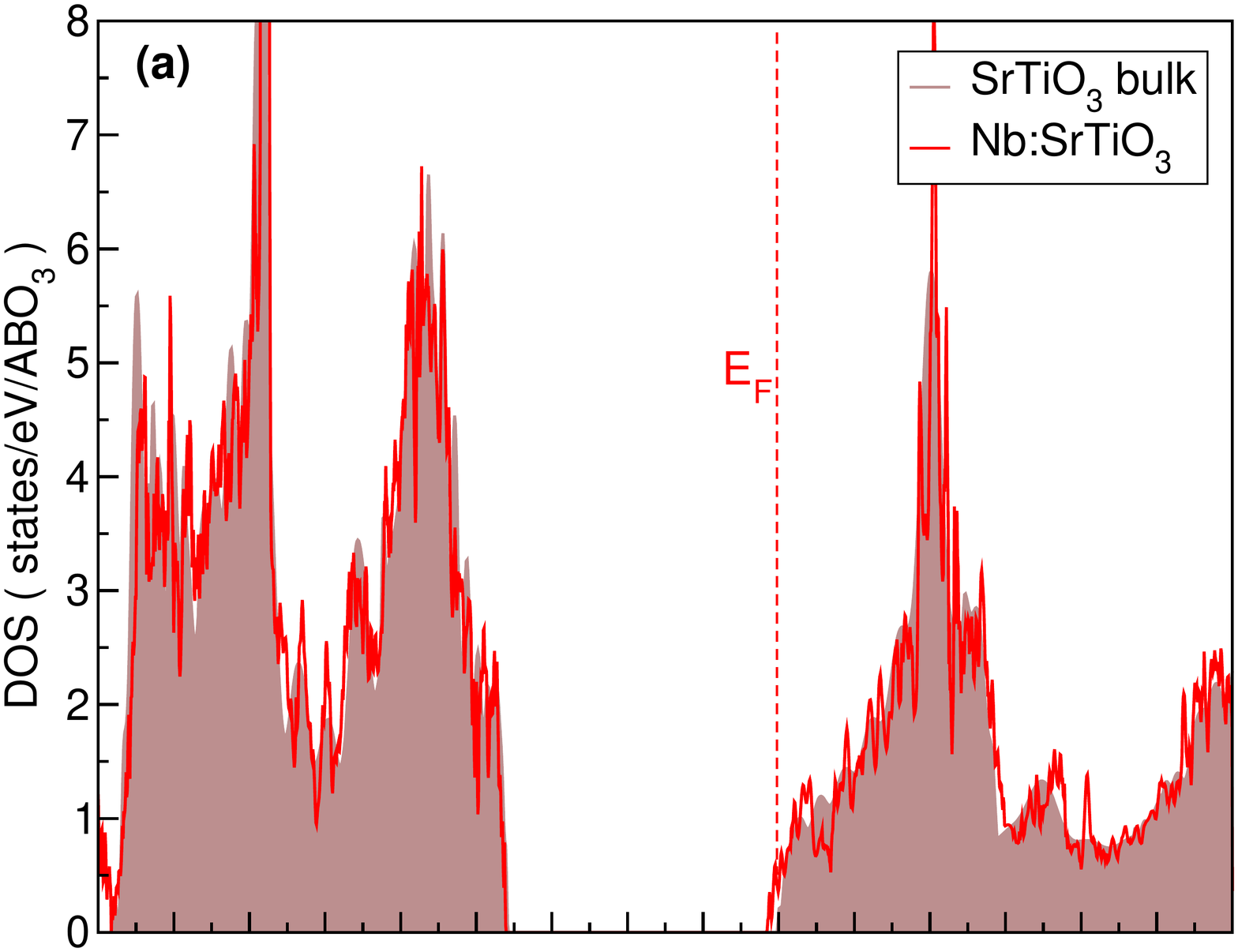}\\[-5pt]
\centering\includegraphics[scale=0.23]{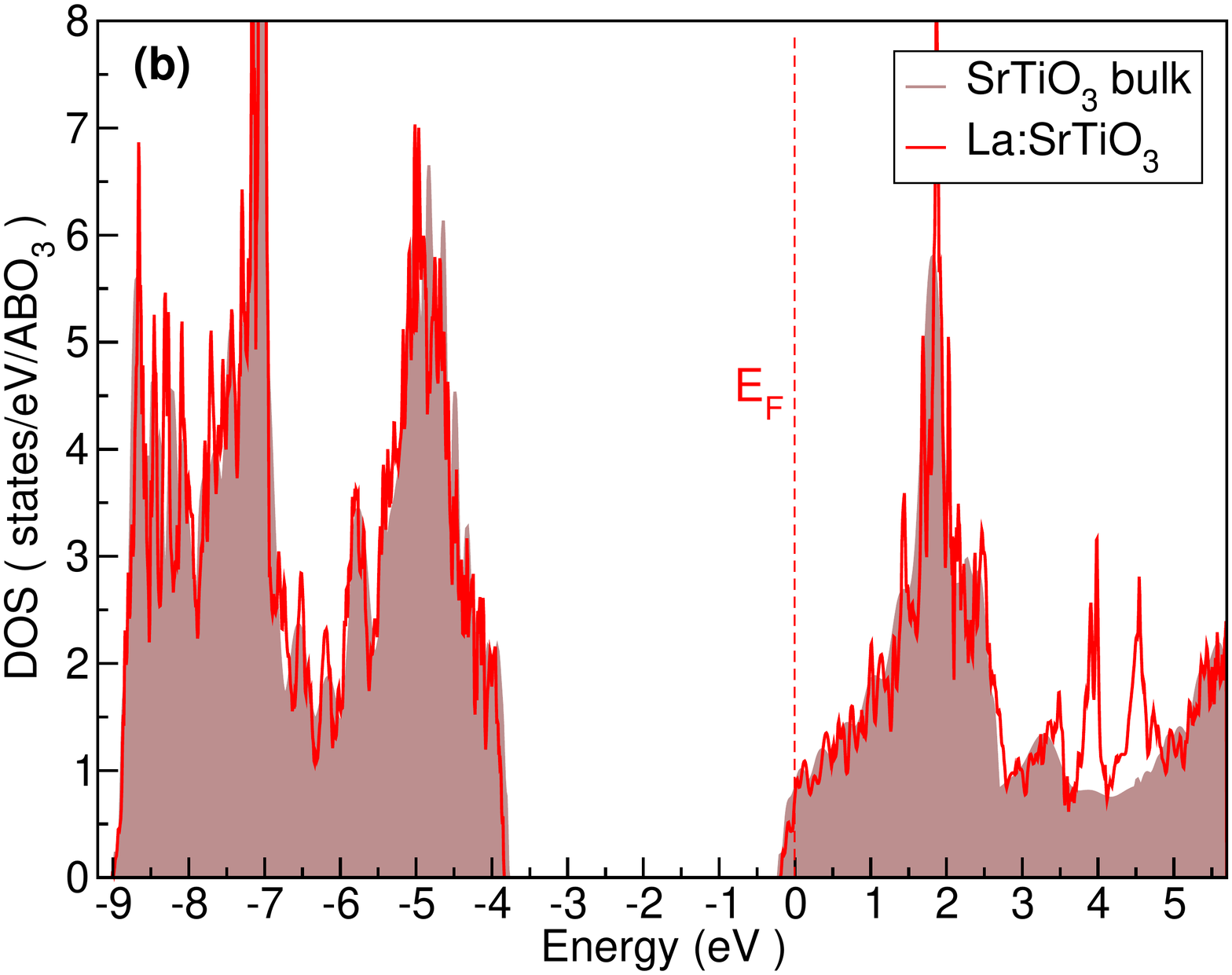}\\[-10pt]
\caption{\label{DOS333STO} (Color online) Density of states (DOS) of: (a) Nb doped \STO\ (Nb:\STO), and (b) La doped \STO\ (La:\STO) at electronic concentrations $n \sim 1.2\times10^{21} cm^{-3}$. The Fermi energy E$_F$ is shown in red dashed line. DOS of bulk \STO\ is shown in background brown color.}
\end{figure}

The promising TE properties of  \STO\  based oxides were found for strong n-type doping ($n \sim10^{21}$ cm$^{-3}$, carrier concentrations which are more than one order of magnitude higher than those of typical TE materials such as PbTe, Bi$_2$Te$_3$). At these high concentrations,  the underestimation of $PF$ within all considered functionals may be generated by the incomplete validity of rigid band structure approximation (Fig.~\ref{PFAllFunc}) . In order to check this, we have considered 3$\times$3$\times$3 \STO\ supercells which explicitly incorporate La and Nb doping elements, and studied their structural, electronic and transport properties. These doping elements introduce one electron to the systems, and change also the electronic states close to the Fermi level (chemical potential) with respect to that of bulk \STO\ (see the density of states DOS from Fig.~\ref{DOS333STO}(a) and (b)). Indeed at these high electronic concentrations the rigid band structure approximation is not completely valid. 

\begin{figure}[t]
\centering\includegraphics[scale=0.25]{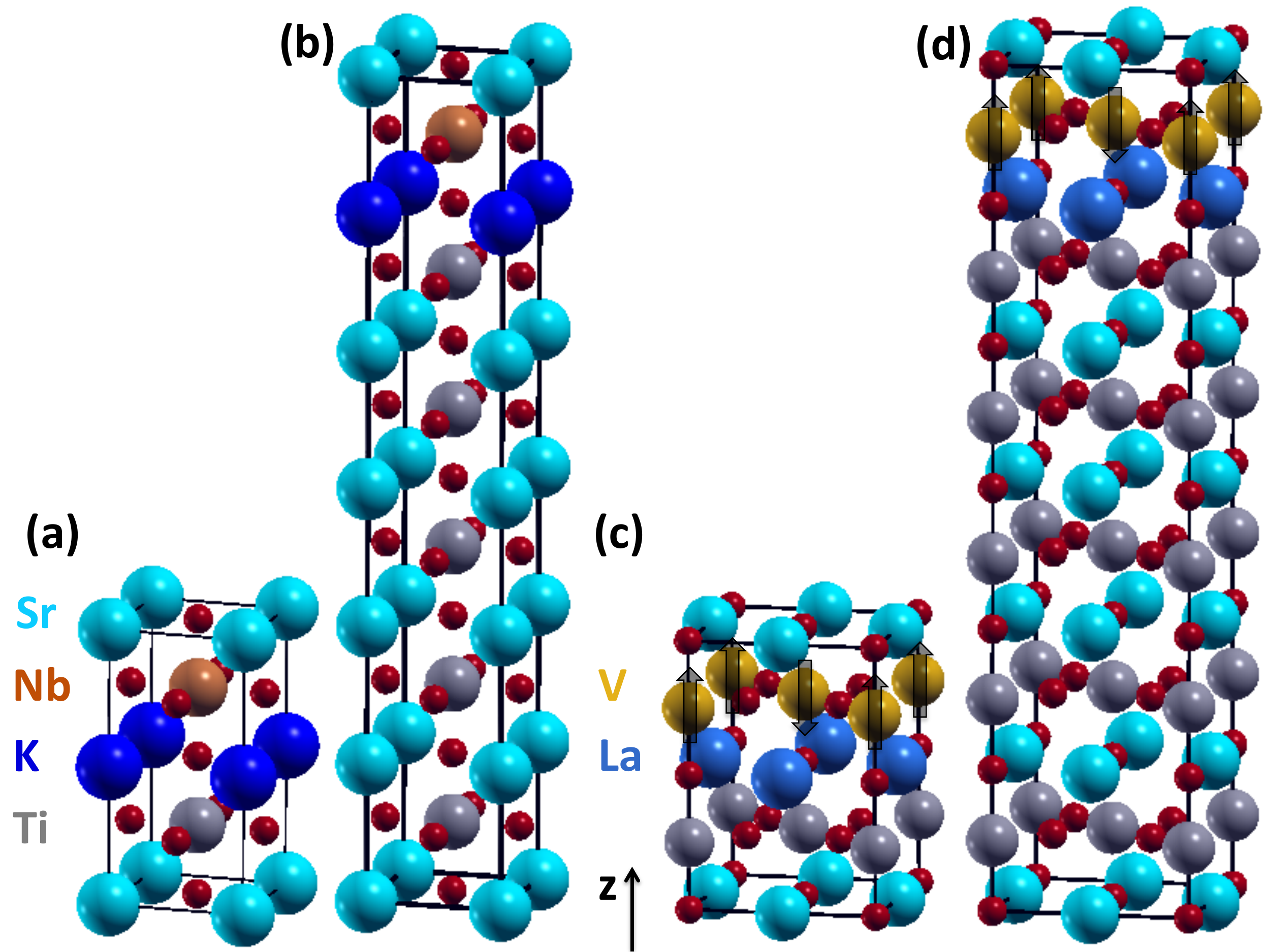}\\[-10pt]
\caption{\label{StrucSTOLVO} (Color online)   (a)-(b) (\STO)$_m$(\KNO)$_1$ nanostructures with nonmagnetic ground state  (m=1,5), and (c)-(d) (\STO)$_m$(\LVO)$_1$ nanostructures with AFM ground state (m=1,5). The AFM spin order on V atoms is shown by arrows.}
\end{figure}

We studied TE properties of 3$\times$3$\times$3 \STO\ supercells, which include explicitly La doping elements (3$\times$3$\times$3 La:SrTiO3). The relaxation time determined by fitting $\sigma_{exp}$ at $n=1\times10^{21}$ cm$^{-3}$ and 300 K is the same as that of bulk \STO\ ($\tau=0.45\times10^{-14}$ s). Although the electronic states near the Fermi level of 3$\times$3$\times$3 La:\STO\ supercell are slightly different than those of bulk \STO\  at high values of $n=1.2\times10^{21}$ cm$^{-3}$ (Fig.~\ref{DOS333STO}(b)), $PF$ is comparable with that of bulk \STO\ (Fig.~\ref{PFAllFunc}). Therefore, the underestimation of experimental power factors $PF_{exp}$ of $\sim$2-3 mW/mK$^2$~\cite{Okuda, Muta, Ohta2005b} is not due to the change of electronic states close to Fermi level generated by doping.
Kinaci {\it et. al.} also showed that La, Nb and Ta doping do not change significantly the electronic states close to the Fermi level, and TE properties of \STO\ alloys are comparable with those of bulk \STO.\cite{Kinaci} $S$ values for \STO\ alloys are slightly lower, whereas $\sigma$ values are slightly larger than those of bulk \STO.\cite{Kinaci} These results suggest that $PF$ of \STO\ alloys is not expected to increase significantly with respect to that of bulk \STO.
 
We assign the underestimation of $PF_{exp}$ to the enhancement of carrier effective mass due to the electron-phonon coupling interaction, which is compatible with the fact that the electronic transport in n-type \STO\ has a polaronic nature.\cite{Mazin} A factor of 3 larger inertial effective mass $m_{i}^*$ was obtained from experimental optical conductivity relative to the theoretical $m_{i}^*$ value of $\sim$0.63$m_e$ estimated within LDA.\cite{Mazin}  At a given carrier concentration, larger experimental effective masses generate larger $S_{exp}$ by lowering the chemical potential relative to CB bottom.
We have estimated $m_{i}^*$ according to the relation~\cite{Snyder}:
\begin{equation}
\frac{1}{m_{i}^*}=\frac{1}{3}(\frac{2}{m_l} + \frac{1}{m_h})
\end{equation}
0.40$m_e$(0.39$m_e$), 6.09$m_e$(6.1$m_e$), and 0.58$m_e$(0.57$m_e$) where $m_l$ and $m_h$ are the light and heavy effective masses 
\begin{widetext}
\begin{figure}[b]
\begin{minipage}[t]{1.0\textwidth}
  \centering\includegraphics[scale=0.32]{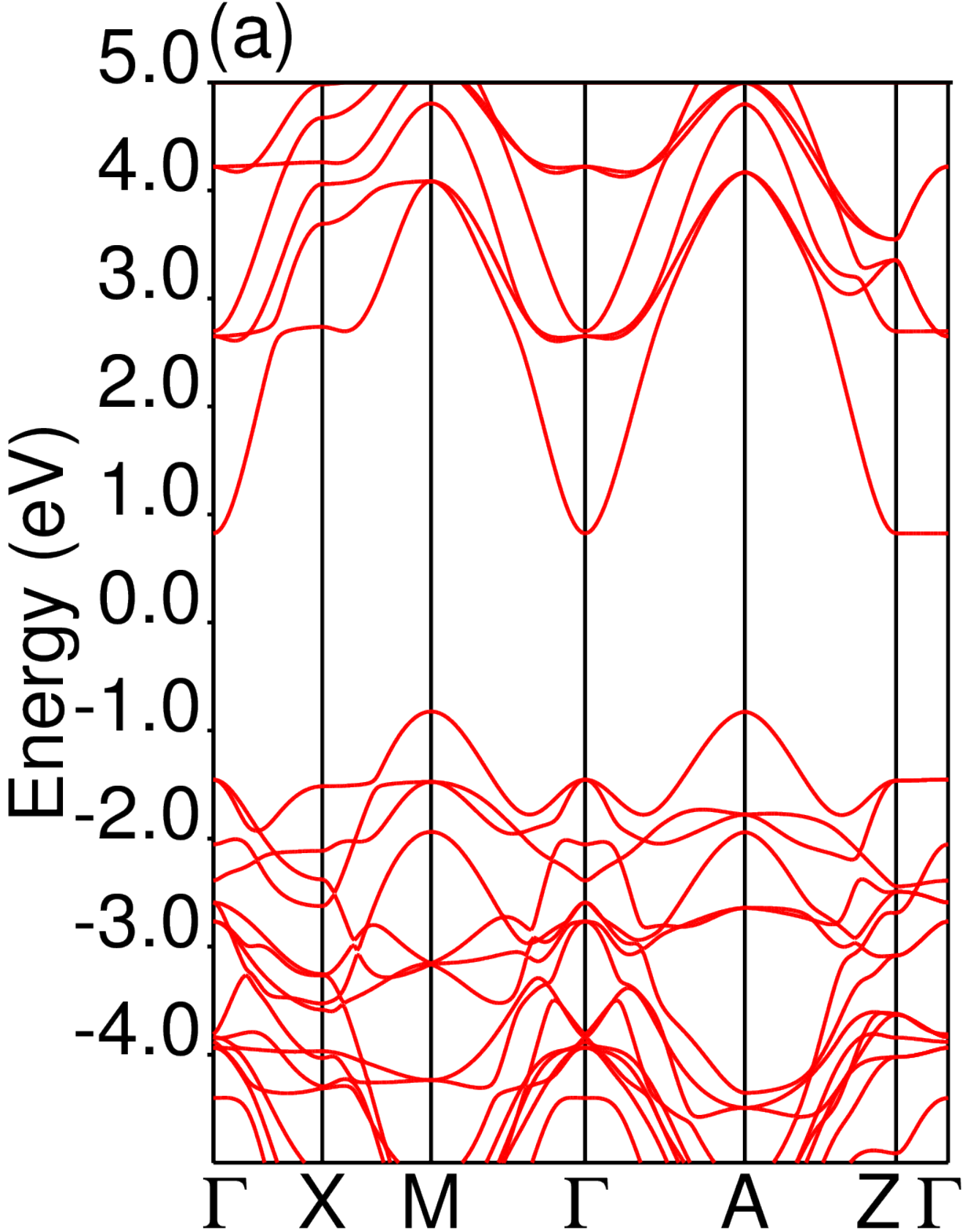}%
  \centering\includegraphics[scale=0.32]{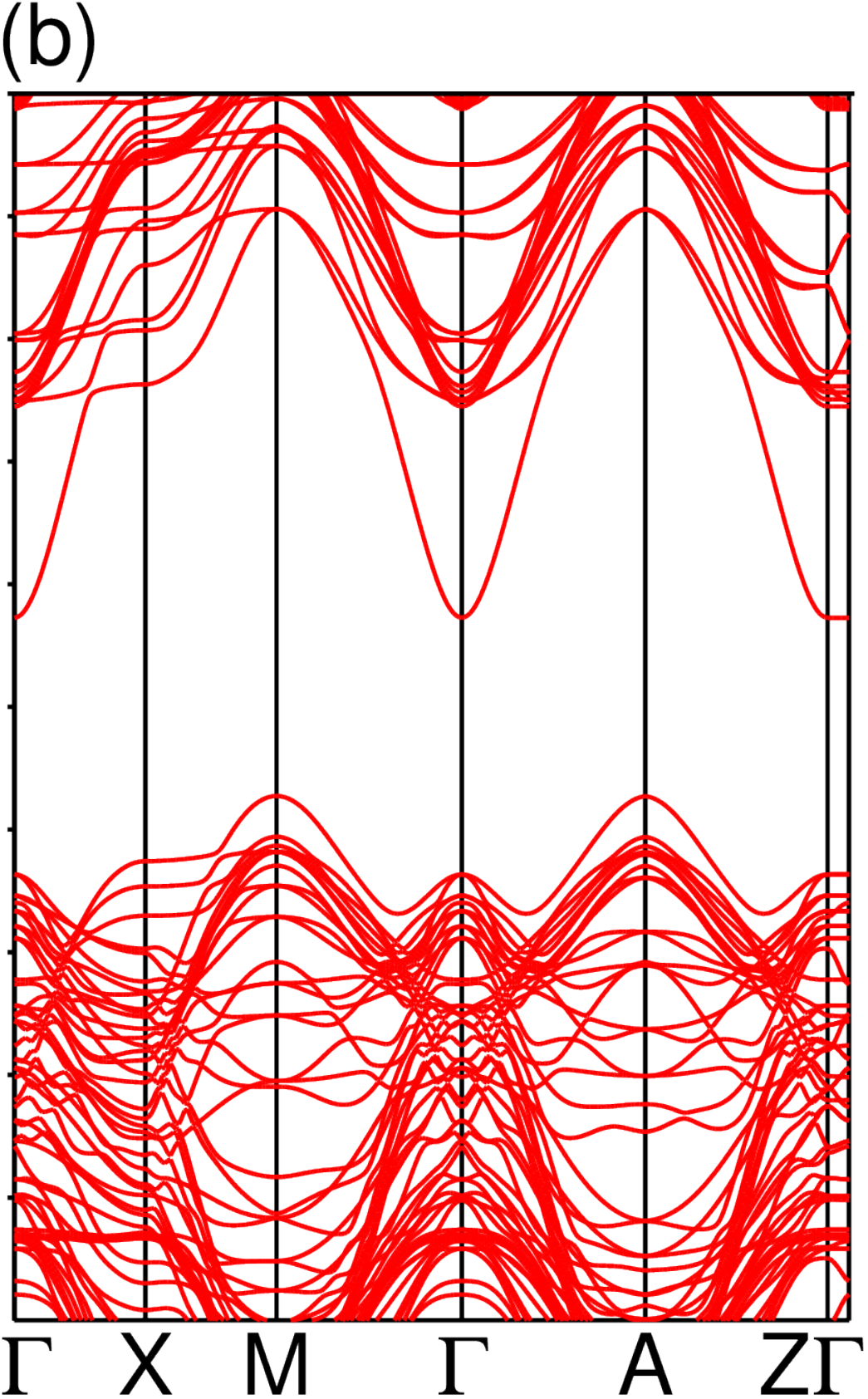}%
  \centering\includegraphics[scale=0.32]{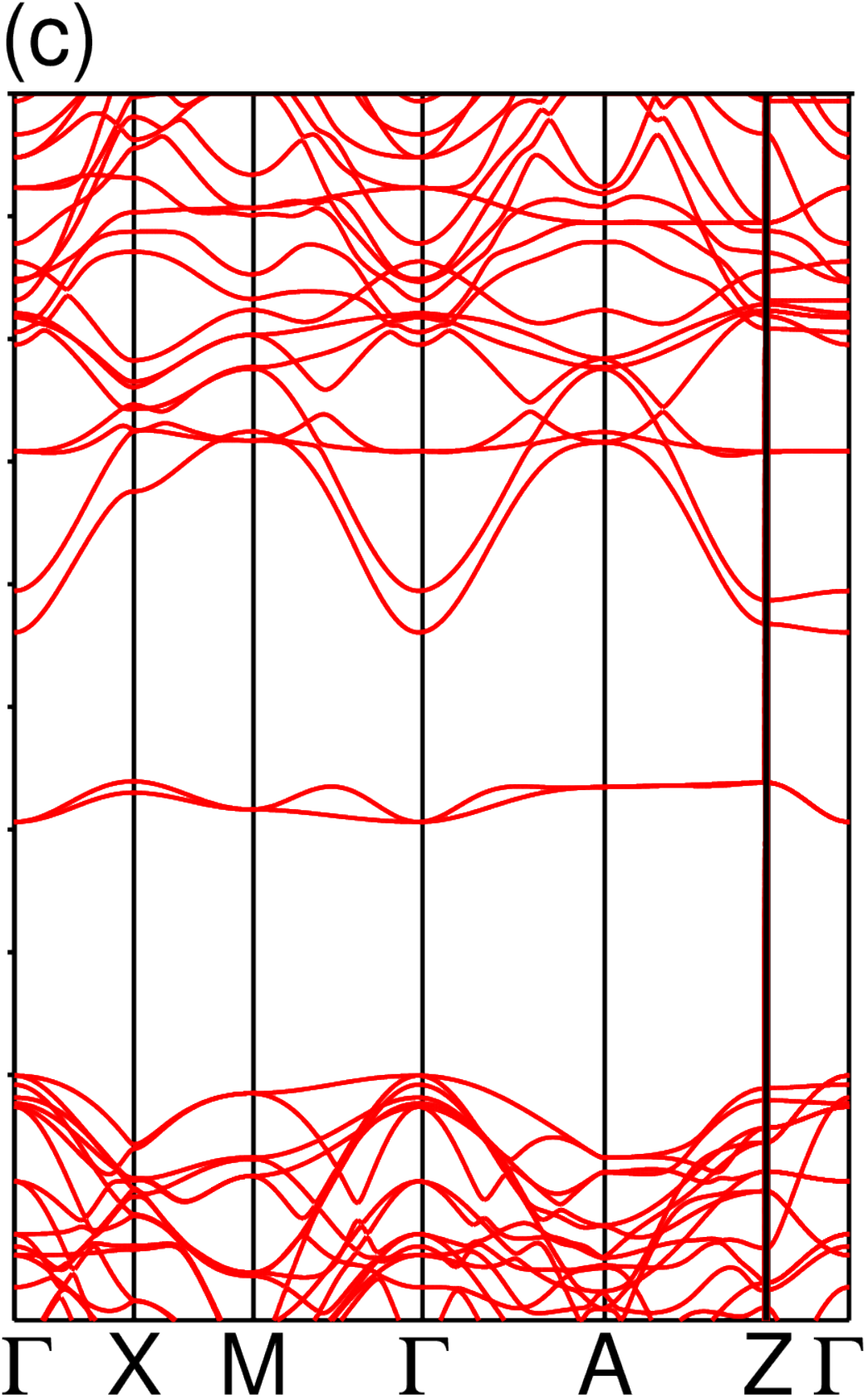}%
  \centering\includegraphics[scale=0.32]{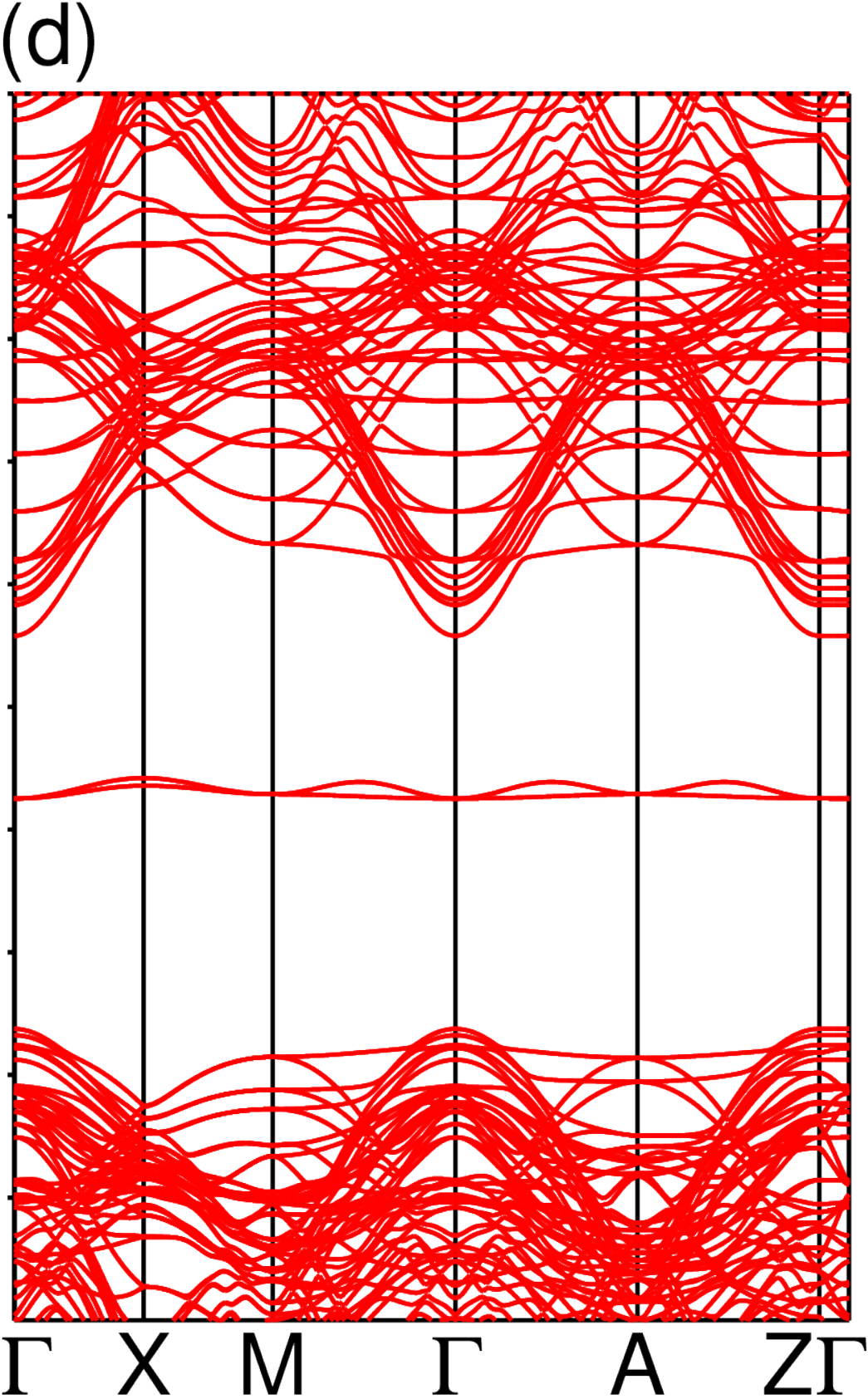}%
  \centering\includegraphics[scale=0.32]{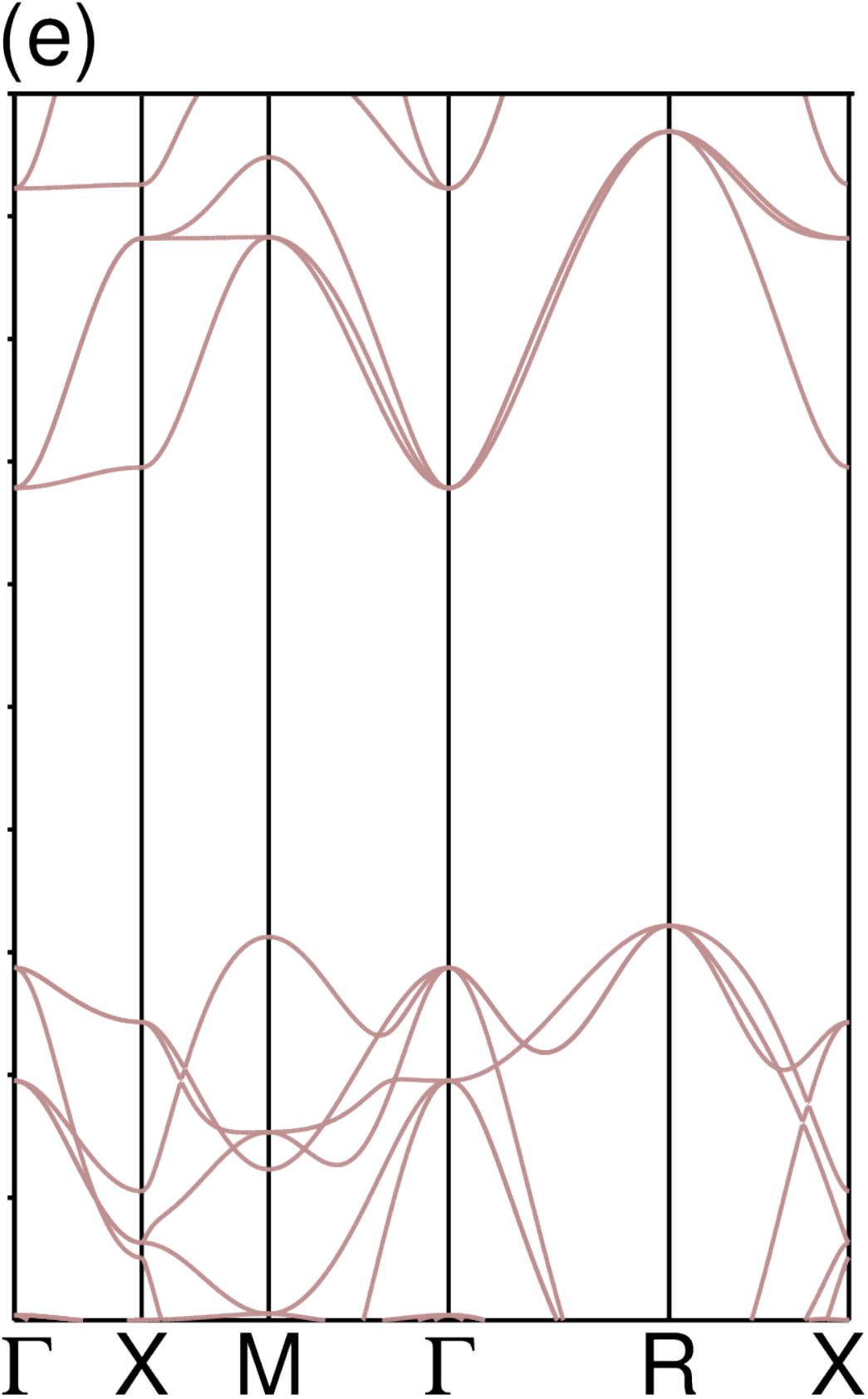}\\[-10pt]
  \caption{\label{BndSTOLVO} Electronic band structure of: (a) (\STO)$_1$(\KNO)$_1$, (b) (\STO)$_5$(\KNO)$_1$, (c) (\STO)$_1$(\LVO)$_1$, (d) (\STO)$_5$(\LVO)$_1$ superlattices, and (e) bulk \STO\ estimated within B1-WC.}
\end{minipage}
\end{figure}
\end{widetext}
of the three fold degenerate Ti $t_{2g}$ bands which form the CB bottom. The estimated $m_l$, $m_h$, and $m_{i}^*$ values are within B1-WC(LDA), respectively.  Since in the transport calculations we do not account for the polaronic nature of \STO\ conductivity, this translate into small values of the estimated relaxation time (see Table~\ref{Table2}). 
 
 \subsection{Band structure engineering in \STO\ based nanostructures}

\begin{figure}[b]
\centering\includegraphics[scale=0.35]{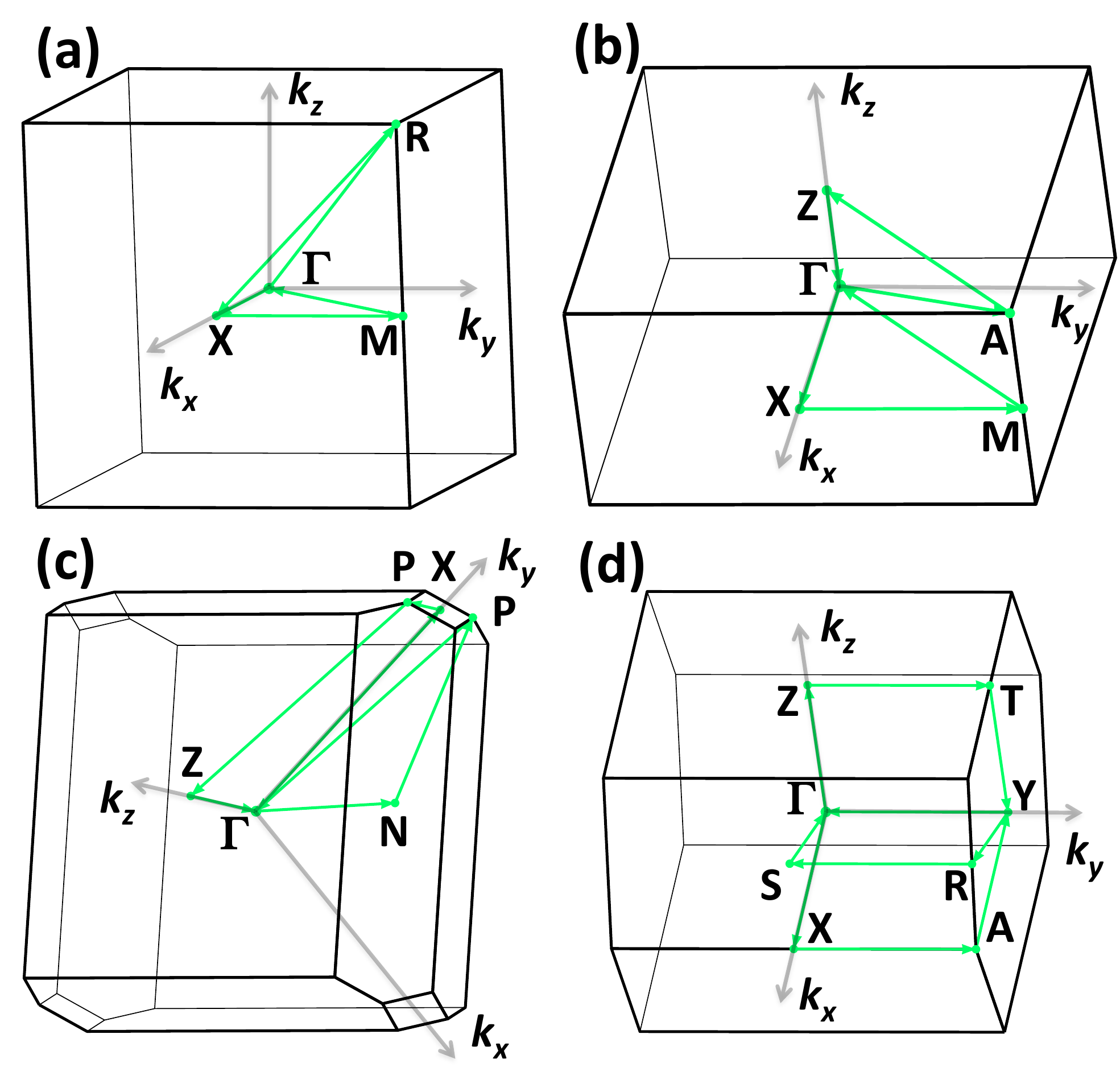}\\[-10pt]
\caption{\label{BrillouinZone} (Color online) Brillouin zone of: (a) simple cubic \STO, (b) tetragonal (\STO)$_m$(\KNO)$_1$ and (\STO)$_m$(\LVO)$_1$ SL (m=1,5), (c) body centered tetragonal SrO[SrTiO$_3$]$_m$ (m=1, 2),  and (d) one face centered tetragonal Sr$_2$CoO$_3$F (ground state structure from  Fig.~\ref{StrucAOABO3}(c)). The high symmetry points along the directions used in electronic band structures, and the orthogonal reciprocal $k_i$ vectors (i=x,y,z) are also shown.}
\end{figure}

\begin{figure}[b]
\centering\includegraphics[scale=0.22]{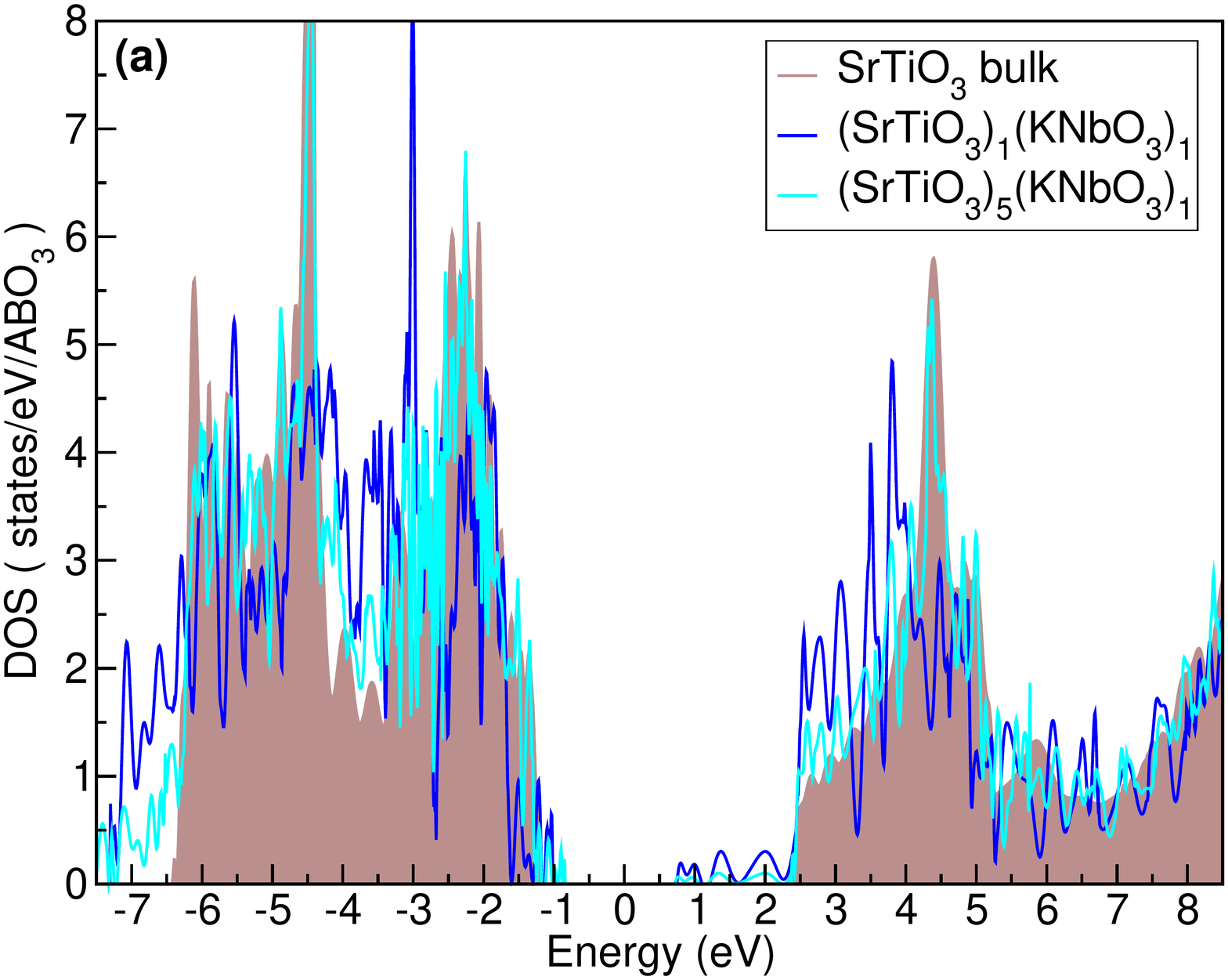}\\[-2pt]
\centering\includegraphics[scale=0.22]{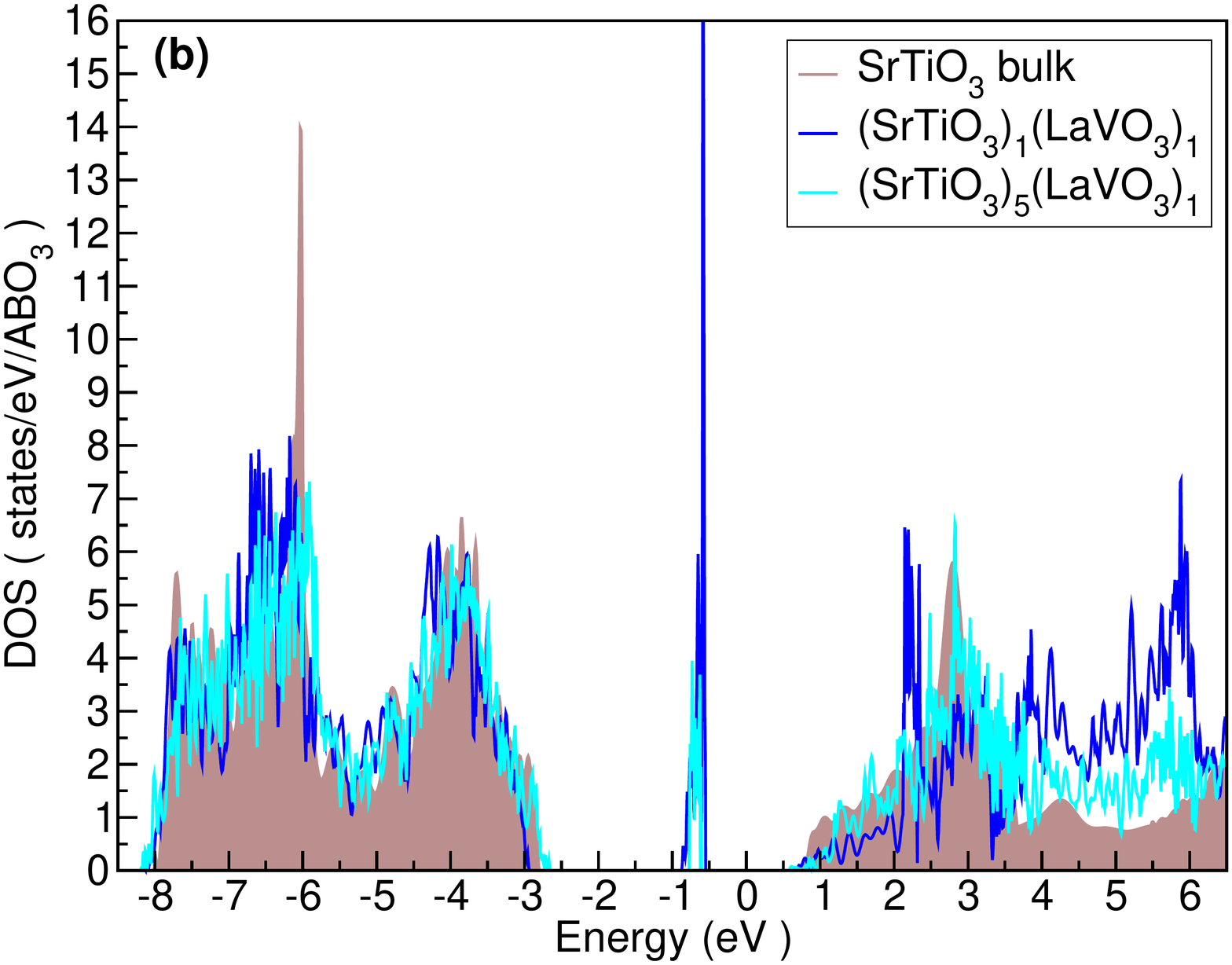}\\[-10pt]
\caption{\label{dosSTOLVO} (Color online) Total density of states (DOS) of: (a) (\STO)$_m$(\KNO)$_1$, and (b) (\STO)$_m$(\LVO)$_1$ superlattices (m=1,5) scaled to ABO$_3$ formula unit. DOS of bulk \STO\ is also shown in background brown color.}
\end{figure}

\begin{figure}[t]
\centering\includegraphics[scale=0.23]{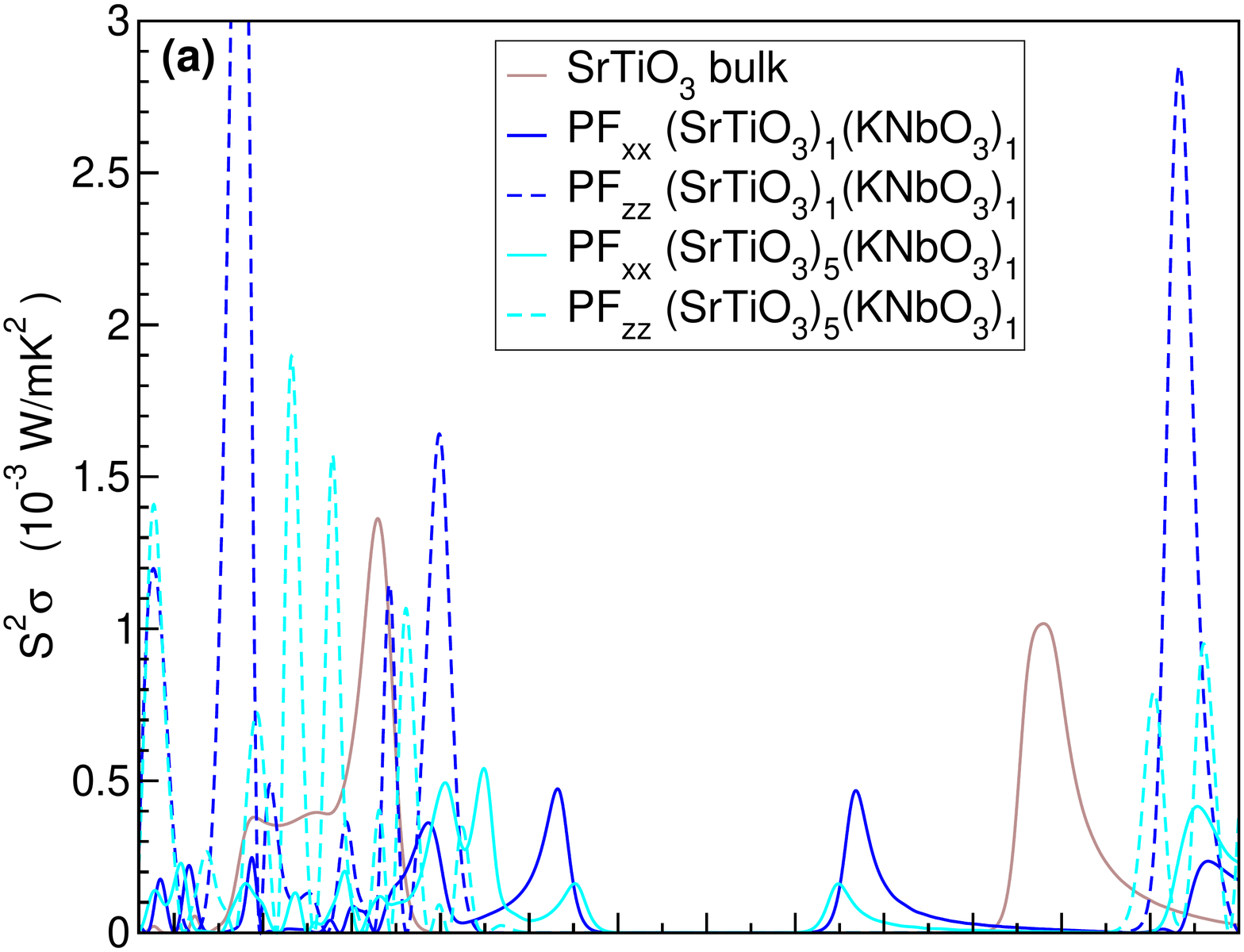}\\[-5pt]
\centering\includegraphics[scale=0.23]{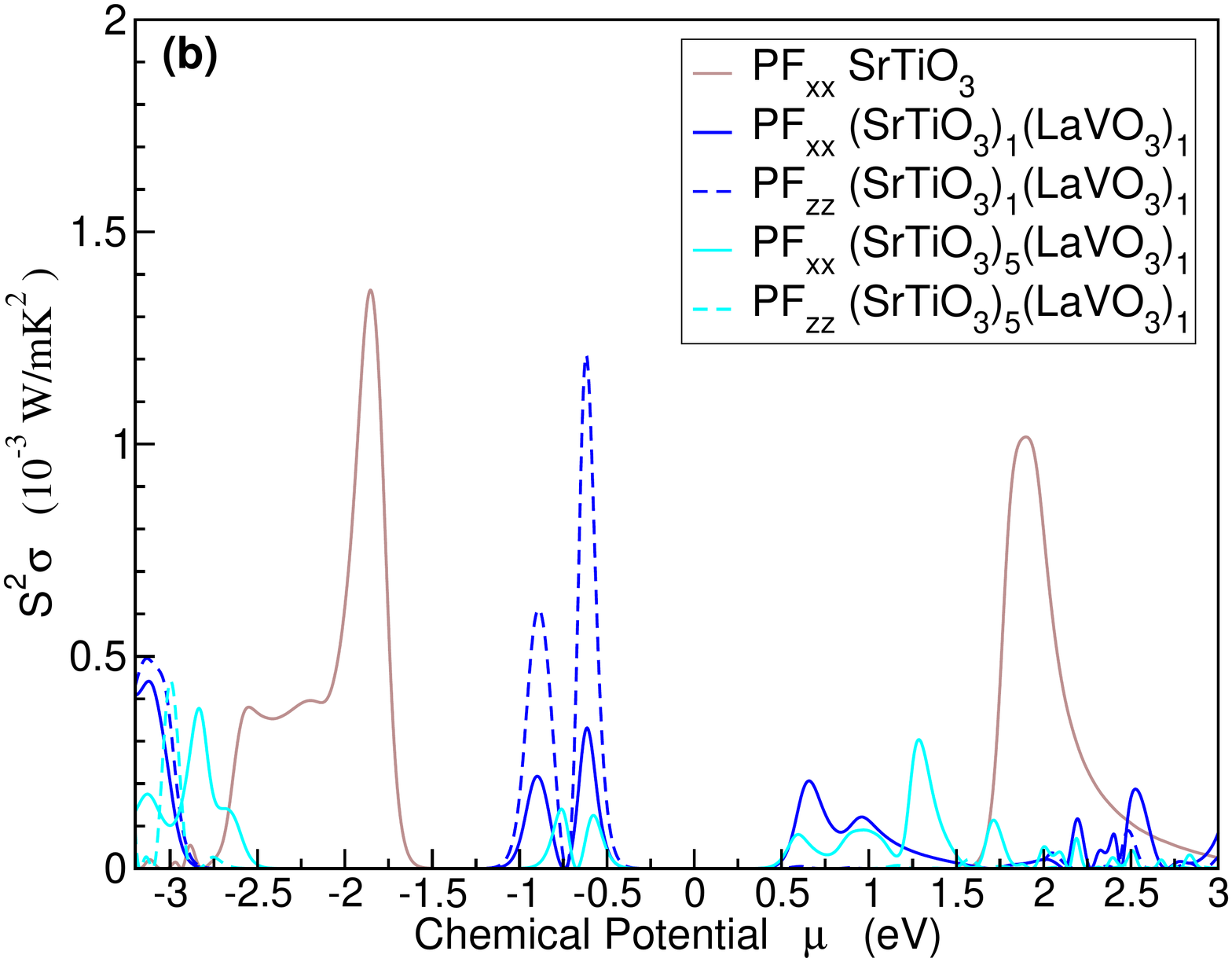}\\[-10pt]
\caption{\label{PFSTOLVO} (Color online) Power factor $PF=S^2 \sigma$ dependence on chemical potential $\mu$ of: (a) (\STO)$_m$(\KNO)$_1$, and (b) (\STO)$_m$(\LVO)$_1$ superlattices (m=1,5) estimated at 300 K within B1-WC using the relaxation time $\tau=0.43\times10^{-14}$ s. }
\end{figure}

 We considered  (\STO)$_m$(\KNO)$_1$ and (\STO)$_m$(\LVO)$_1$ SL nanostructures with m=1, 5 for which we studied the electronic and transport properties, and described the relation between size of nanostructures and their TE properties by looking at the effect of quantum confinement on $PF$. (\STO)$_m$(\KNO)$_1$ SL with m=1, 5 have a nonmagnetic ground state. Their structures and electronic band structures are shown in Figs.~\ref{StrucSTOLVO}(a),(b), and~\ref{BndSTOLVO}(a),(b). In comparison to bulk \STO, (\STO)$_1$(\KNO)$_1$ SL possess smaller $E_g$, and an electronic band which is very flat along $\Gamma$Z direction and dispersive in the other  
orthogonal directions of the Brillouin zone (see Fig.~\ref{BrillouinZone}(b)). This very flat-and-dispersive band forms the CB bottom and has a Nb $d_{xy}$ orbital character. The electronic states associated to this very anisotropic band, which participate in the electronic transport, have a reduced weight (short $\Gamma$Z distance in the Brillouin zone). The weight is proportional with the density of states DOS and the carrier pocket volumes inside of the Brillouin zone. The reduced weight can be seen more easily from DOS scaled to ABO$_3$ formula unit (f.u.) (see Fig.~\ref{dosSTOLVO}(a)). The electronic states inside of \STO\ band gap have a small weight, which generate in the inplane direction power factors $PF_{xx}$ smaller than that of bulk \STO\  (Fig.~\ref{PFSTOLVO}(a)). In the cross plane direction, (\STO)$_1$(\KNO)$_1$ SL show large power factors $PF_{zz}$ but at very high $n$ values (chemical potential $\mu$ $\sim$2.75 eV) which can not be achieved in  experiment. Increasing the quantum confinement in the case of (\STO)$_5$(\KNO)$_1$ SL, decreases the weight of very anisotropic flat-and-dispersive Nb $d$ band. The decrease in weight of this anisotropic band can be seen from the electronic band structure and DOS (see Figs.~\ref{BndSTOLVO}(b),~\ref{dosSTOLVO}(a)), and produces  a $PF$ drop relative to (\STO)$_1$(\KNO)$_1$ SL (Fig.~\ref{PFSTOLVO}(a)).


SL formed by (\STO)$_m$(\LVO)$_1$ with m=1, 5 have an antiferromagnetic (AFM) ground state. Their structures and electronic band structures are shown in Figs.~\ref{StrucSTOLVO}(c),(d), and~\ref{BndSTOLVO}(c),(d). These SL possess two electronic bands laying inside of \STO\ band gap, which are very flat in $\Gamma$A and AZ directions, and weakly dispersive in the other directions of Brillouin zone (Fig.~\ref{BrillouinZone}(b)). These flat bands create a narrow energy distribution with a very large weight on the top of valence band, being generated by V $d_{xz}$  and $d_{yz}$ orbitals. The very large weight of this narrow energy distribution can be seen from DOS, and this distribution generates $PF$s smaller than those of \STO\ (see Figs.~\ref{dosSTOLVO}(b), and~\ref{PFSTOLVO}(b)). This shows that a single or multiple very flat bands having large effective masses in all directions of Brillouin zone are not able to enhance TE performance, since the charge carriers associated to such flat bands are very localized and unable to participate in electronic transport.
Increasing the quantum confinement in (\STO)$_5$(\LVO)$_1$ SL, also lowers the weight of narrow energy distribution and $PF$ of these SL (see Figs.~\ref{dosSTOLVO}(b), and~\ref{PFSTOLVO}(b)).

\begin{figure}[b]
\centering\includegraphics[scale=0.25]{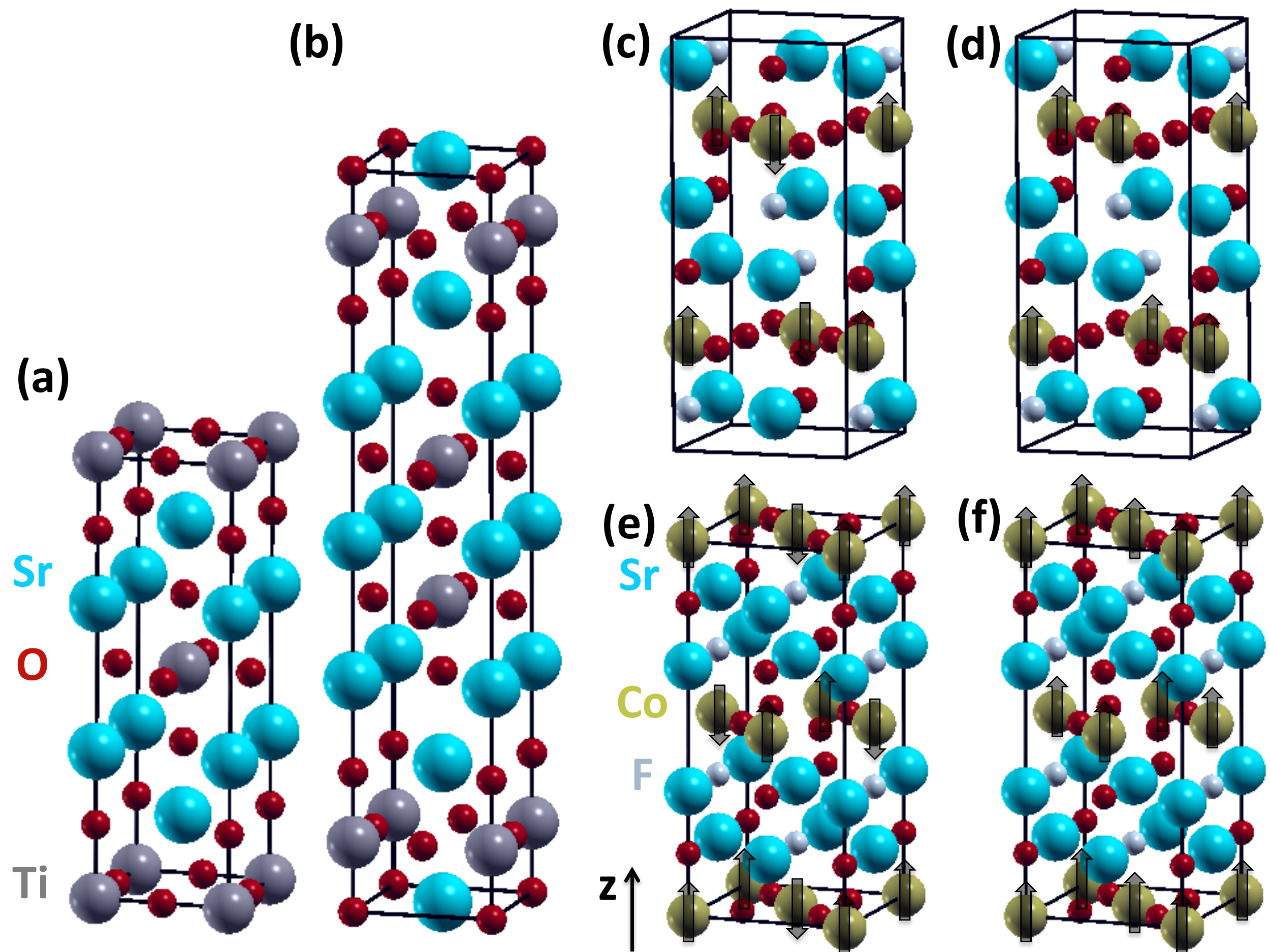}\\[-10pt]
\caption{\label{StrucAOABO3} (Color online)   Structures of (a) Sr$_2$TiO$_4$, and (b) Sr$_3$Ti$_2$O$_7$. Model structures of  Sr$_2$CoO$_3$F with: (c)
AFM order on Co and one F atom in apical position, (d) FM order on Co and one F atom in apical position, (e) AFM order on Co and two F atoms in apical position, and (f) FM order on Co and two F atoms in apical position. The spin order on Co atoms is shown by arrows.}
\end{figure}

 \subsection{Band structure engineering in AO[ABO$_3$]$_m$ naturally-ordered Ruddlesden-Popper phases}

\begin{widetext}
\begin{figure}[h]
\begin{minipage}[h]{1.0\textwidth}
  \centering\includegraphics[scale=0.32]{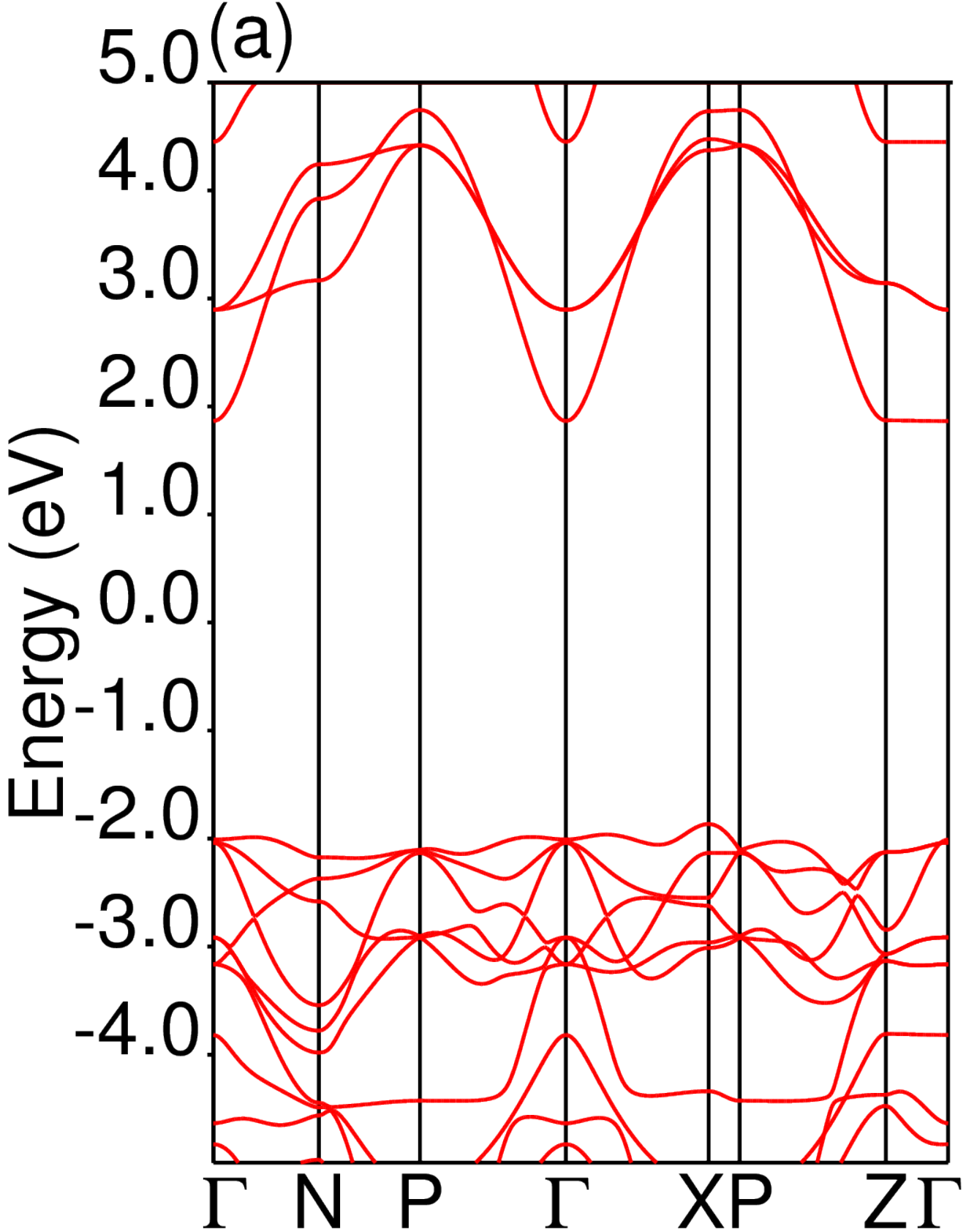}%
  \centering\includegraphics[scale=0.32]{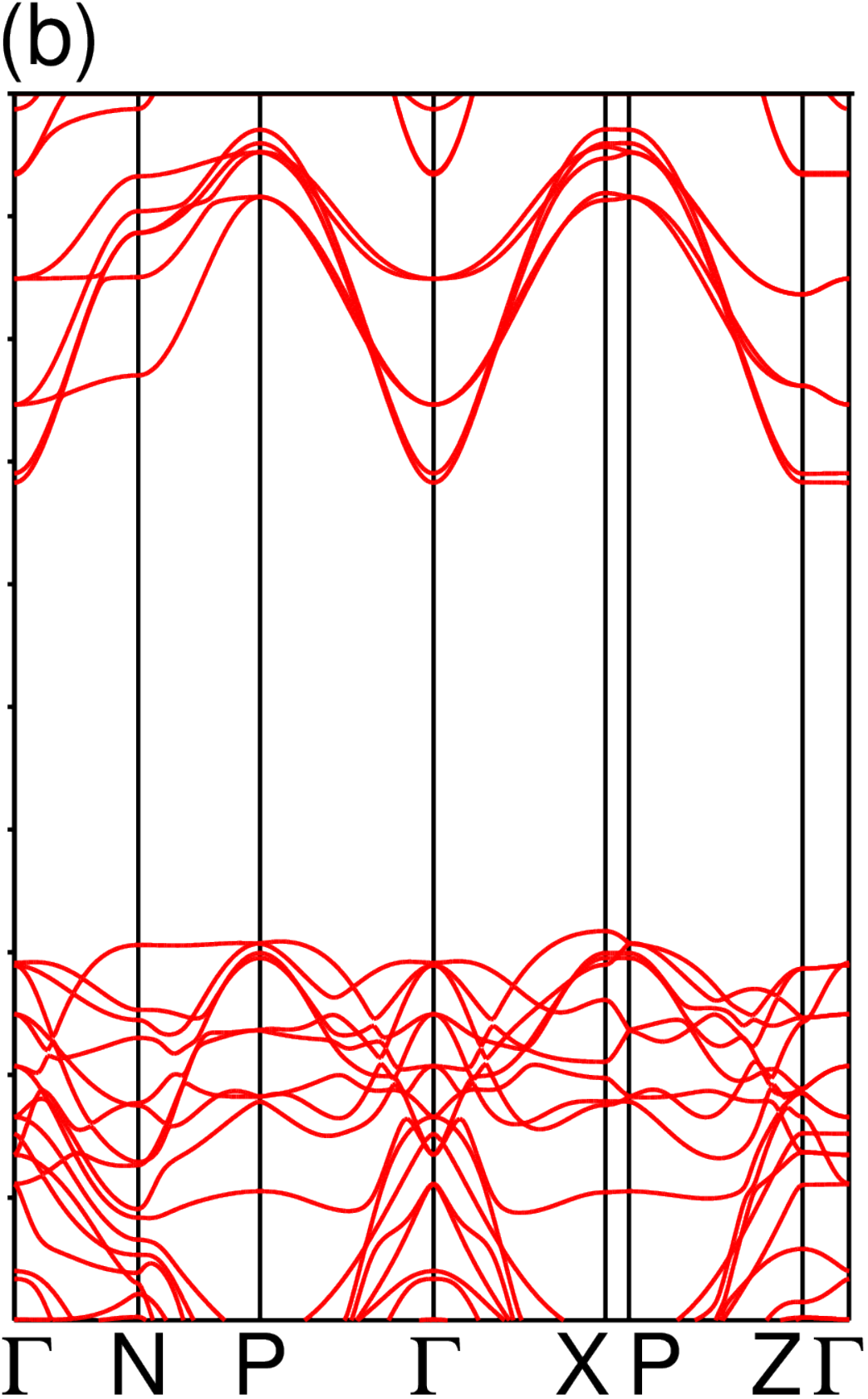}%
  \centering\includegraphics[scale=0.32]{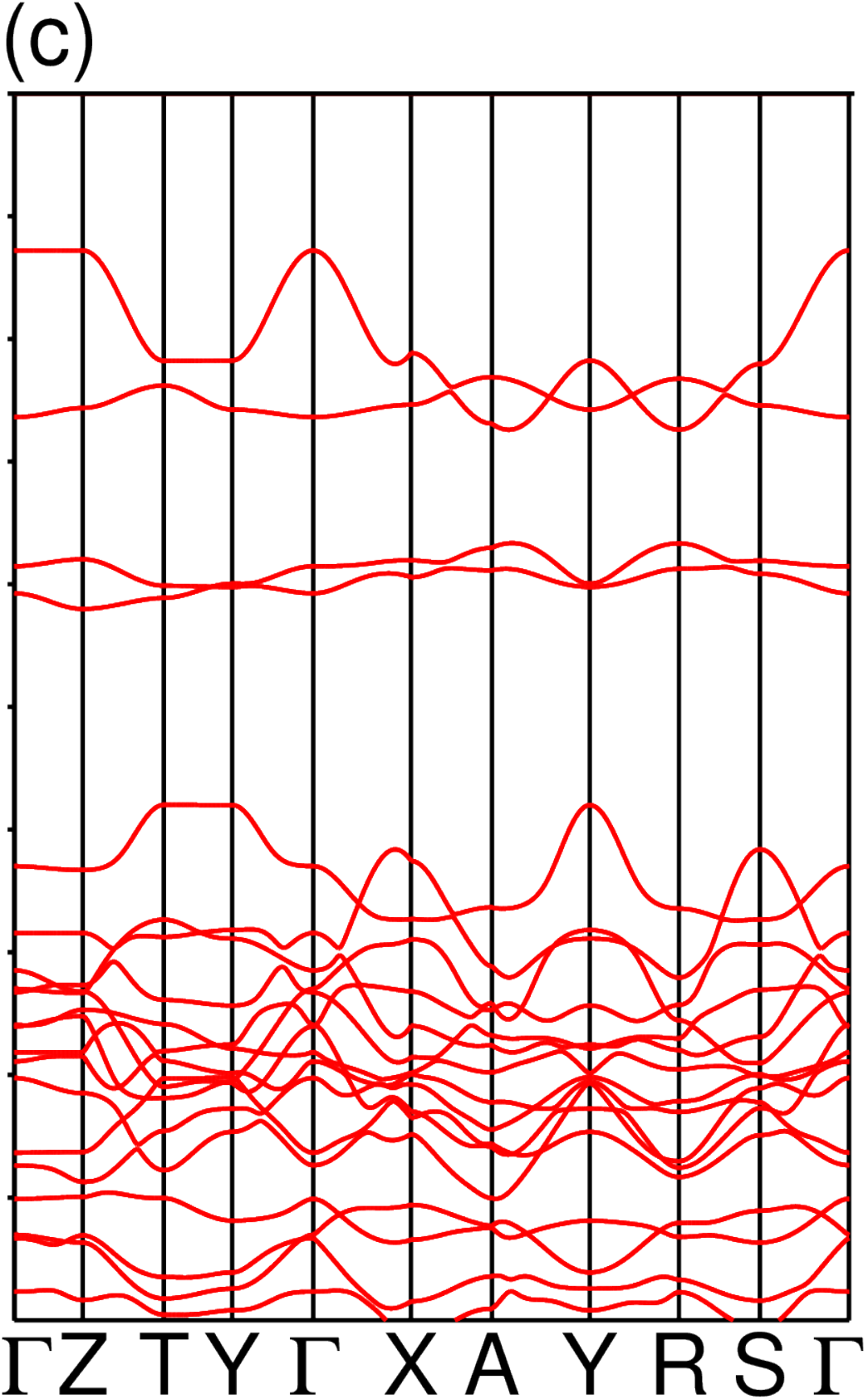}%
  \centering\includegraphics[scale=0.32]{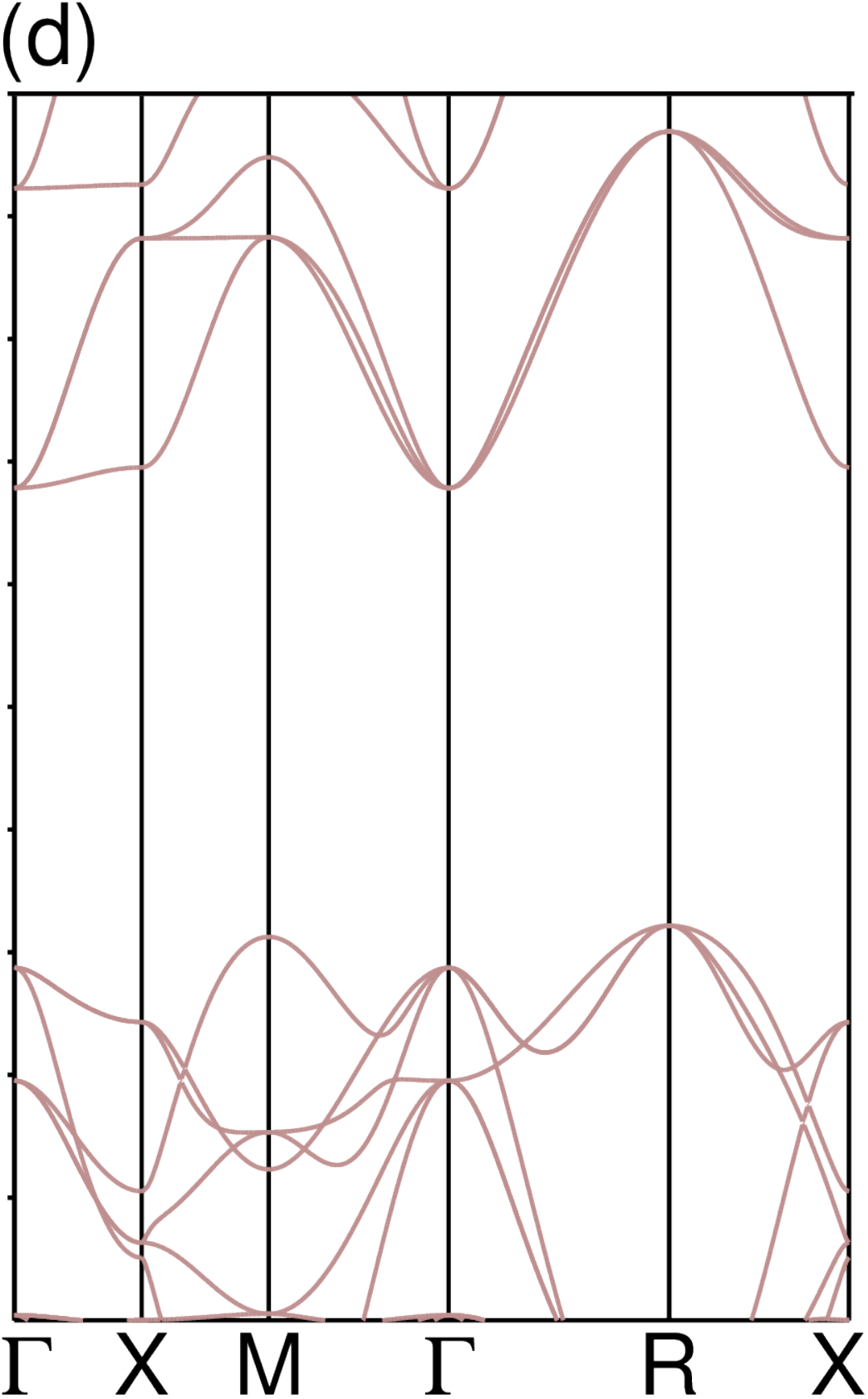}\\[-10pt]
  \caption{\label{BndAOABO3} Electronic band structure of: (a) Sr$_2$TiO$_4$, (b) Sr$_3$Ti$_2$O$_7$, (c) Sr$_2$CoO$_3$F (ground state structure from  Fig.~\ref{StrucAOABO3}(c)), and (d)  bulk \STO\ estimated within B1-WC.}
\end{minipage}
\end{figure}
\end{widetext}

\begin{figure}[b]
\centering\includegraphics[scale=0.25]{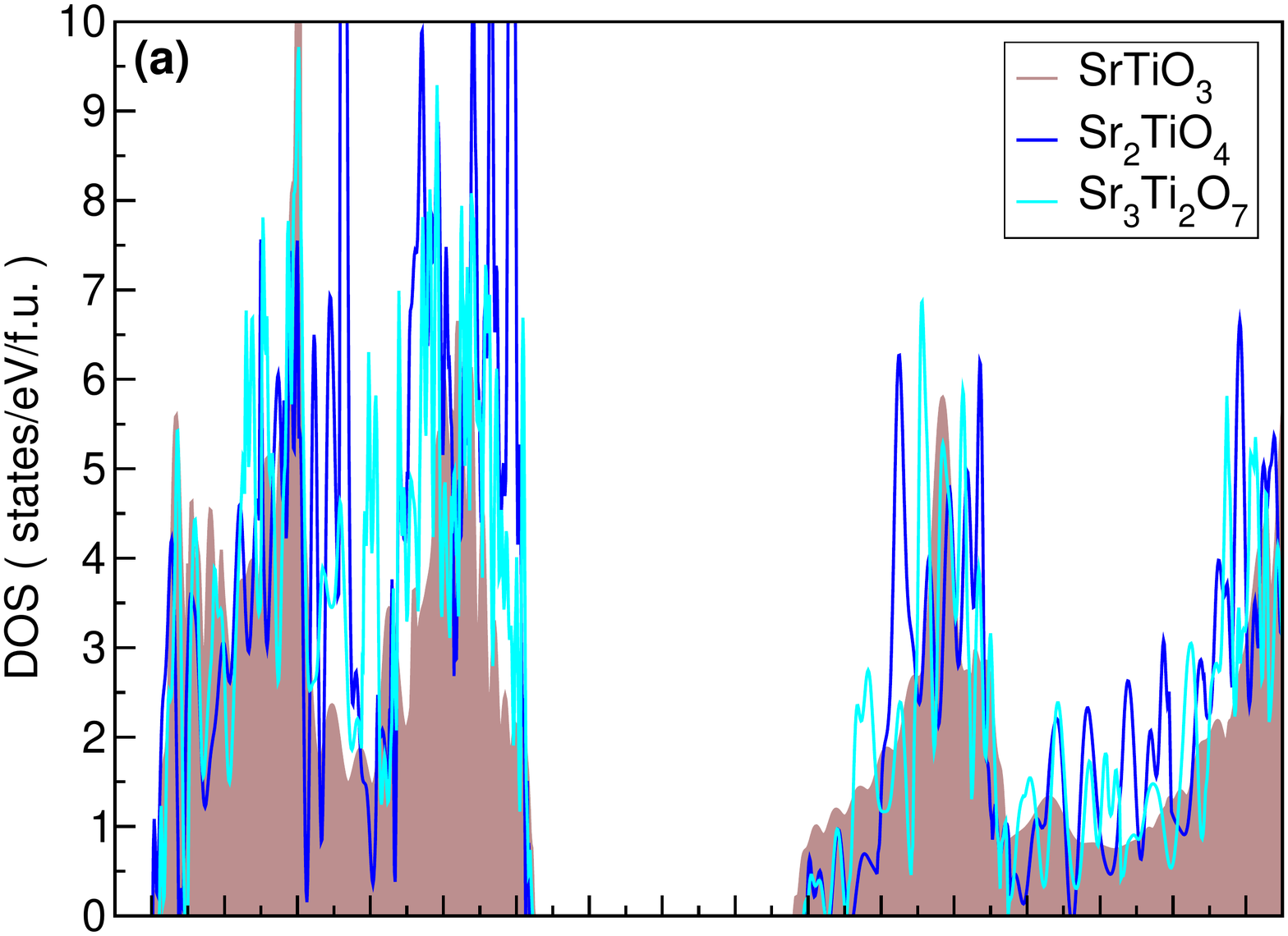}\\[-12pt]
\centering\includegraphics[scale=0.25]{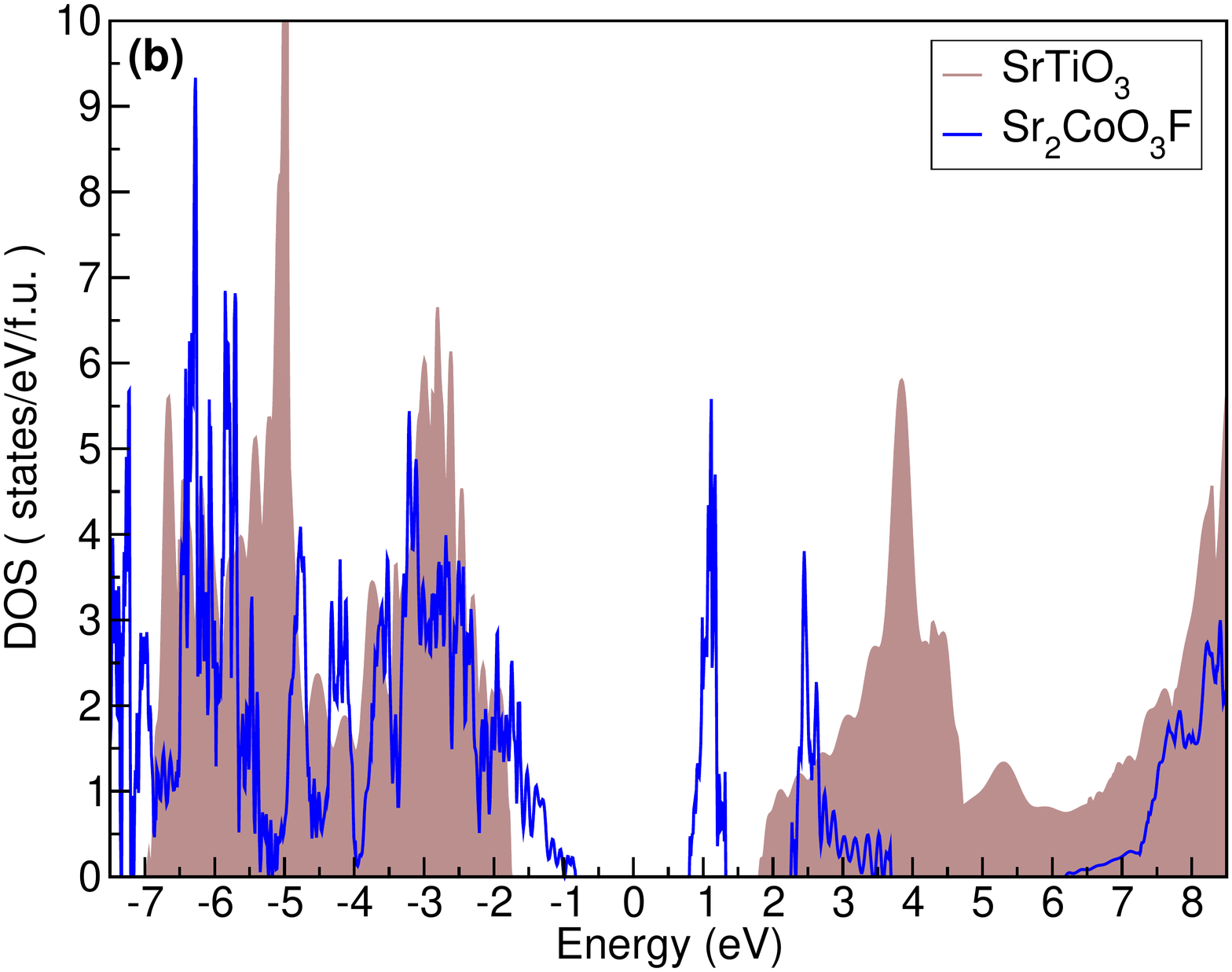}\\[-10pt]
\caption{\label{dosAOABO3} (Color online) Total density of states (DOS) of: (a) Sr$_2$TiO$_4$ and Sr$_3$Ti$_2$O$_7$,  and (b) Sr$_2$CoO$_3$F (ground state structure from  Fig.~\ref{StrucAOABO3}(c))  scaled to formula unit (f.u.= Sr$_2$TiO$_4$, Sr$_{1.5}$TiO$_{3.5}$, and Sr$_2$CoO$_3$F, respectively). }
\end{figure}

 Highly anisotropic flat-and-dispersive bands can be found also in AO[ABO$_3$]$_m$ Ruddlesden-Popper naturally-ordered compounds. These compounds are formed from ABO$_3$ perovskite layers separated by an AO atomic layer  and can nowadays be grown epitaxially with atomic-scale control.\cite{Schlom}  To search for highly anisotropic bands, we have considered SrO[SrTiO$_3$]$_m$ (m=1 and 2) and SrO[SrCoO$_2$F]$_1$  compounds (Fig.~\ref{StrucAOABO3}).  
 The insertion of SrO atomic layer in the crystallographic direction $Oz$ creates the quantum confinement of electronic states in $\Gamma$Z direction from the band structure of Sr$_2$TiO$_4$ and Sr$_3$Ti$_2$O$_7$ (see Fig.~\ref{BndAOABO3}(a),(b)). 
It can be seen that CB bottom is formed by such very anisotropic bands, which generate narrow energy distributions with small weights (small length of $\Gamma$Z direction). The small weights of these distributions close to CB minimum (energy $\sim$1.75 - 2 eV) can be seen more easily from scaled DOS per f.u. of 
Sr$_2$TiO$_4$ and Sr$_3$Ti$_2$O$_7$, (Fig.~\ref{dosAOABO3}(a)).

\begin{figure}[t]
\centering\includegraphics[scale=0.25]{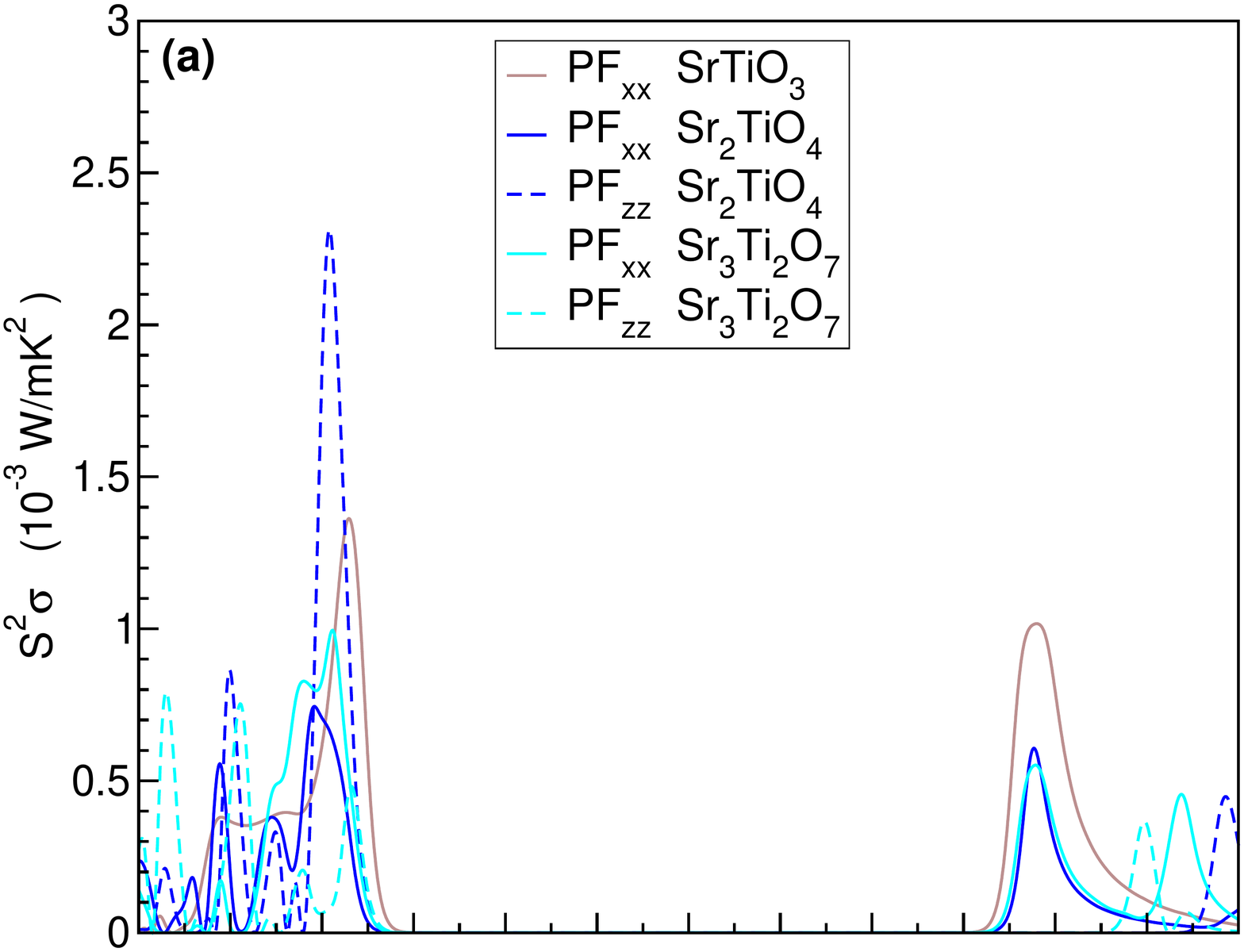}\\[-5pt]
\centering\includegraphics[scale=0.25]{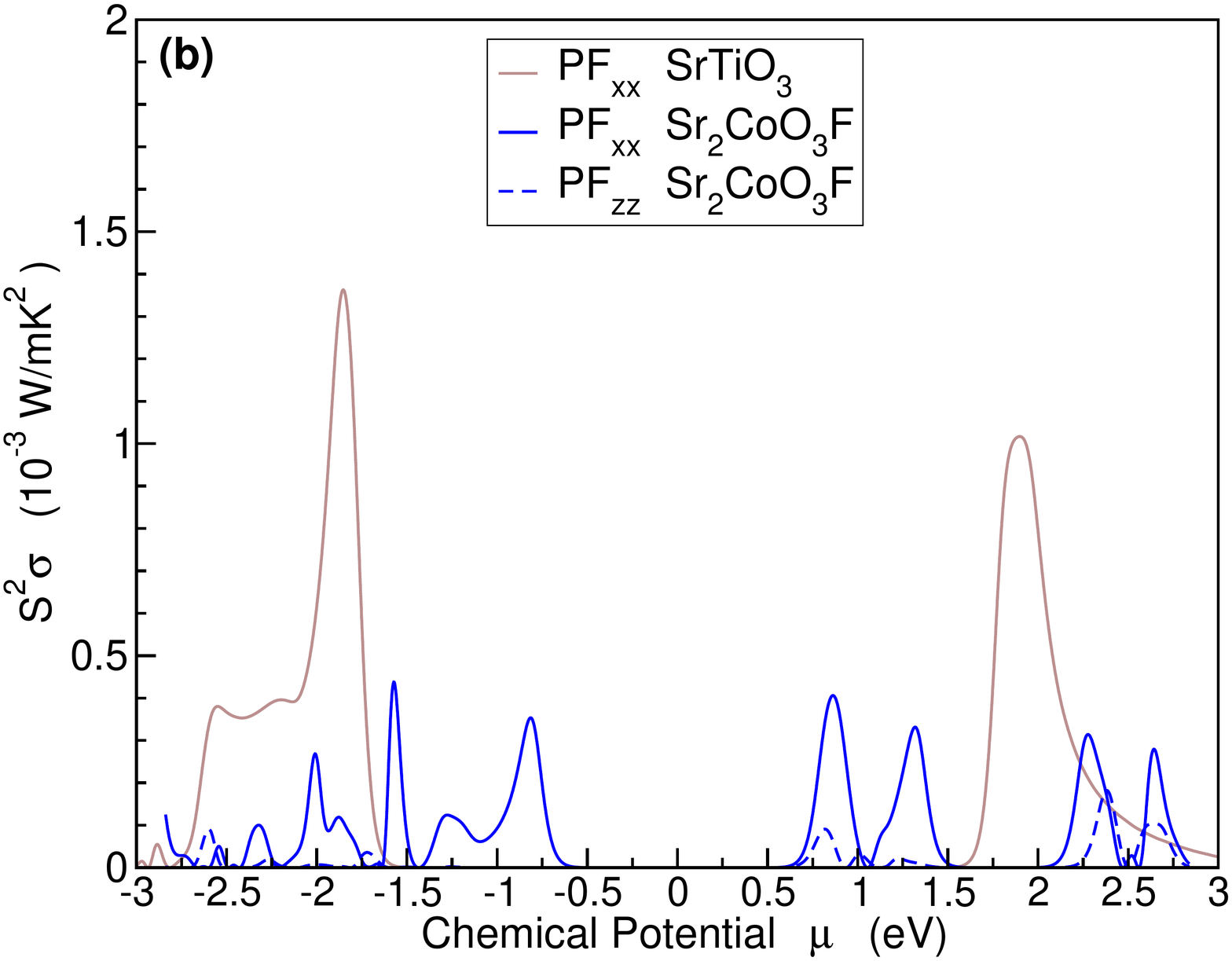}\\[-10pt]
\caption{\label{PFAOABO3} (Color online) Power factor $PF=S^2 \sigma$ dependence on chemical potential $\mu$ of: (a) Sr$_2$TiO$_4$ and Sr$_3$Ti$_2$O$_7$, and (c) Sr$_2$CoO$_3$F naturally-ordered superlattices estimated at 300 K within B1-WC using the relaxation time $\tau=0.43\times10^{-14}$ s. }
\end{figure}

TE properties have been estimated using the same value of $\tau$ as that of bulk \STO, because we want to compare the electronic contribution given by the electronic band structure of these naturally-ordered compounds to that of bulk \STO. Due to the small weight of narrow energy distribution, the n-type $PF$ corresponding to the chemical potential $\sim$1.75 - 2 eV in $Ox$ direction ($PF_{xx}$) is smaller than that of bulk \STO\ (Fig.~\ref{PFAOABO3}(a)). In the approximation that $\tau$ of SrO[SrTiO$_3$]$_m$ compounds is similar to that of bulk \STO, the n-type $PF$s is not improved. 
 
From Co based AO[ABO$_3$]$_m$ compounds, we have explored the cobalt oxyfluoride Sr$_2$CoO$_3$F, in which F substitute O form apical position of CoO$_6$ octahedron.\cite{Tsujimoto} In Figure~\ref{StrucAOABO3}(c-f) are shown the model structures in which F substitute one O atom (Fig.~\ref{StrucAOABO3}(c),(d)) or two O atoms (Fig.~\ref{StrucAOABO3}(e),(f)) form apical positions and Co atoms have AFM/FM order. The analysis of structural properties shows that the ground state structure is the structure in which F substitute one O atom form apical positions and Co atoms have AFM order (Fig.~\ref{StrucAOABO3}(c)). These results are in agreement with the experimental study, which finds G-type antiferromagnetic order of Co in Sr$_2$CoO$_3$F.\cite{Tsujimoto}  For the ground state structure, we studied the electronic  
and transport properties. The electronic band structure of Sr$_2$CoO$_3$F contains two electronic bands with Co $d$ orbital character in the (0.75eV, 1.25eV) energy interval (Fig.~\ref{BndAOABO3}(c)). These bands do not have a very anisotropic character, requirement identified to maximize $PF$.\cite{Bilc2015}  As a result the energy distribution of the two Co bands is narrow and has large weight, which can be seen from DOS (Fig.~\ref{dosAOABO3}(b)). In the approximation that $\tau$ of Sr$_2$CoO$_3$F is comparable to that of \STO, the transport calculations show that $PF$ of these compounds is smaller than that of \STO\ (Fig.~\ref{PFAOABO3}(b)).

\subsection{Comparison of the different nanostructures with bulk \STO}

In the approximation that $\tau$ of the considered nanostructures is comparable to that of bulk \STO, none of the nanostructures shows higher TE performance than bulk \STO, in spite of the fact that some of them possess highly anisotropic flat-and-dispersive TM $d$ electronic bands. In addition, the electronic states associated to these bands which participate in transport must have significant weights in order to maximize $PF$ and $n$. The weights are proportional to DOS which depends on the density of states effective mass $m_{d}^* = \gamma^{2/3} (m_{l}^{2}m_{h})^{1/3} $, where $\gamma$ is the carrier pocket degeneracy or band multiplicity.\cite{Bilc2006}  Therefore in Table~\ref{Table3}, we show  the estimated values of DOS($\mu$)/f.u. at the chemical potential $\mu$ which optimizes n- and p-type $PF$s, $m_{d}^*$,  and band anisotropy ratio $R =m_{h}/m_{l}$. The usual anisotropic behavior is for heavy masses $m_h$ across SL direction ($Oz$), and light masses $m_l$ in the inplane SL direction ($Ox$, $Oy$). 
Although (\STO)$_1$(\KNO)$_1$ SL show very large $R$ and large $m_{d}^*$ values, their DOS($\mu$)/f.u. corresponding to the maximum n-type $PF$ is about one order of magnitude smaller than that of \STO. 
The confinement of Nb $d_{xy}$ states achieved in these SL is able to create very large anisotropic electronic bands, but with small weights of the electronic states participating in transport  due to their small carrier pocket volume in the Brillouin zone. (\STO)$_1$(\LVO)$_1$ SL at Z point of VB maximum, show unusual band anisotropic behavior with $m_l$ along $z$ direction and $m_h$ along $x$ and $y$ directions, which gives large values for $R$, $m_{d}^*$, DOS($\mu$)/f.u. at the p-type $PF$ maximum, and larger p-type power factor across SL direction ($PF_{zz}$) than along SL direction ($PF_{xx}$). Similar to (\STO)$_1$(\KNO)$_1$ SL, the confinement of Ti $t_{2g}$ ($d_{xy}$) states participating in transport of Sr$_2$TiO$_4$ and Sr$_3$Ti$_2$O$_7$ naturally-ordered SL creates large anisotropy ratios, but at the same time detrimental reduced weights. On the other hand, Co $t_{2g}$ ($d_{xz}$, and $d_{yz}$) states involved in the n-type transport of Sr$_2$CoO$_3$F have large weights, but small anisotropy ratio. For comparison, we show in Table~\ref{Table3} the corresponding values for the full Heusler Fe$_2$TiSi which shows very large n-type $PF$s.\cite{Bilc2015}  These very large $PF$s are achieved for concomitant large anisotropy ratio and weights, and small $m_l$ effective mass which gives large carrier mobilities along the transport direction.

\begin{table}[t]
\caption{\label{Table3} Heavy $m_h$($m_e$) and light $m_l$($m_e$) effective masses, anisotropy ratio $R$, carrier pocket degeneracy (or band multiplicity) $\gamma$, and density of states effective mass $m_{d}^*$($m_e$) for CB minima and VB maxima of  \STO\  and related superlattices, where $m_e$ is free electron mass. DOS($\mu$)/f.u. (states/eV*f.u.) corresponding to $\mu$ which maximizes the n- or p-type $PF$s, where f.u. is ABO$_3$,  A$_2$BO$_4$,  or A$_{1.5}$BO$_{3.5}$ unit cell with only one TM atom per cell. The values for Fe$_2$TiSi are included for comparison.  }
 \begin{tabular*}{0.5\textwidth}%
    {@{\extracolsep{\fill}}cccccccc}
\hline\hline
                                           &                            & $m_h$ & $m_l$ & $R$     & $\gamma$ & $m_{d}^*$&DOS($\mu$)  \\ 
\hline
\STO                                  &CB($\Gamma$)& 6.1       & 	0.4      &   15.3      &	    3               &	2.06         &  14.6    \\
(\STO)$_1$(\KNO)$_1$&CB($\Gamma$)& 219.4   & 	0.27    &   812.6   &	    1               &	2.52         &  1.5    \\
(\STO)$_1$(\LVO)$_1$&CB($\Gamma$)& 4.66      &	         0.41    &   11.4     &	    1               &	0.92         &  2.4    \\
                                           &VB(Z)                  & 15.62   &	  0.8    &   19.5    &	    1               &	5.8         &  90.9    \\
                                           &VB(X)                  & 13.92   &        1.73    &       8.1   &         1               &	3.47       &              \\
Sr$_2$TiO$_4$              &CB($\Gamma$)& 74        &	        0.42     &   176.2      &	    1               &	2.36         &  4.7    \\ 
Sr$_3$Ti$_2$O$_7$     &CB($\Gamma$)& 50        &  	0.4      &   125      &	    1               &	2.0         &  4.9    \\ 
Sr$_2$CoO$_3$F          &CB(Z)                  & 5.7       &	1.54    &   3.7      &	    1               &	2.91         &  32    \\
                                           &                             &              &         2.8      &  2.0        &                          &                     &           \\
                                           &VB(T)                   & 244.6       &	0.36    &   679.4      &	    1               &	3.65         &  5    \\
                                           &                             &                   &    0.55    &  444.7       &                     &                     &           \\
Fe$_2$TiSi$^a$             &CB($\Gamma$)& 90              &	0.2       &   450         &	    3               &	3.19         &  31.2    \\ 
\hline\hline
\multicolumn{8}{l}{ $^a$From Ref.~\cite{Bilc2015}.  }\\
\end{tabular*}
\end{table}

\section{Conclusions}

Using the concept of electronic band structure engineering we tried to design materials possessing highly anisotropic electronic bands in (\STO)$_m$(\KNO)$_1$ and (\STO)$_m$(\LVO)$_1$ (m=1 and 5) artificial superlattices, and in SrO[SrTiO$_3$]$_m$ (m=1 and 2) and SrO[SrCoO$_2$F]$_1$ naturally-ordered superlattices. In spite of the fact that almost all superlattices possess such highly anisotropic electronic bands created by the confinement of TM $d$ states, which is a signature of low-DET, their $PF$s are not better than that of \STO. The origin of this TE performance is the small weights of electronic states participating in transport, which are associated to the highly anisotropic electronic bands. The experimental evidences for the decreased effective TE performance of quantum wells and two-dimensional electron gas systems, caused by the contribution of barrier layers used to create the confinement, support our conclusion.\cite{Hicks, Ohta2007} Another detrimental effect on TE performance is the large $m_l$ values along the transport direction of \STO\ and related oxide materials, which are a factor $\sim$2 larger than those of Fe based Heusler compounds~\cite{Bilc2015}, and a factor of $\sim$16  larger than those of usual thermoelectrics such as PbTe.\cite{Bilc2006}   If we account for the polaronic conductivity of  \STO, these factors are $\sim$ 3 times larger. Although \STO\ possesses highly directional TM $d$ electronic states active in transport, these states do not generate very large PF's. The origin of this TE performance is the low electron mobility as a result of the polaronic nature of electrical conductivity. In \STO\ and related perovskite oxides, there is an important TM $d$ - O $p$ hybridization with covalent character, which appears to favour the polaronic conductivity. Binary TM oxides possessing high structural symmetries with stronger TM $d$ - $d$ atomic interactions, may show high anisotropy, large weights and high mobilities of the charge carriers giving improved thermoelectric performance over ABO$_3$ perovskite oxides.

\begin{acknowledgments}

The authors acknowledge financial support from the Romanian National Authority for Scientific Research, CNCS-UEFISCDI, Project number PN-II-PT-PCCA-2013-4-1119. Ph. G. acknowledges the ARC project AIMED and the F.R.S.-FNRS project HiT4FiT.

\end{acknowledgments}

\end{document}